\documentclass[12pt]{article}

\usepackage{times}
\usepackage{comment}
\usepackage{xcolor}
\usepackage{empheq}
\usepackage{amsmath}
\usepackage{amssymb}
\usepackage{ulem,xpatch}
\usepackage{multirow}
\usepackage{lineno}
\usepackage[section]{placeins}

\newcommand{\hide}[1]{\relax}

\newcommand{\Og}{\ensuremath{\Omega}}
\newcommand{\Om}{\ensuremath{\Omega_\mathrm{m}}}

\newcommand{\Gm}{\ensuremath{\Gamma_\mathrm{m}}}
\newcommand{\Gdec}{\ensuremath{\Gamma_\mathrm{dec}}}

\newcommand{\meff}{m_\mathrm{eff}}

\newcommand{\chim}{\chi_\mathrm{m}}

\newcommand{\vcrn}[1]{\mathrm{g_\mathrm{0}}_{#1}}

\newcommand{\bnth}{\ensuremath{\bar n_\mathrm{th}}}

\newcommand{\kB}{\ensuremath{k_\mathrm{B}}}

\newcommand{\Gmeas}{\ensuremath{\Gamma_\mathrm{meas}}}

\newcommand{\xzpf}{\ensuremath{x_\mathrm{zpf}}}

\newcommand{\Fth}{\ensuremath{\hat{F}_\mathrm{th}}}

\newcommand{\an}[1]{\hat{a}_{#1}}
\newcommand{\andag}[1]{\hat{a}^\dagger_{#1}}

\newcommand{\Detn}[1]{\Delta_{#1}}
\newcommand{\kk}[1]{\kappa_{#1}}
\newcommand{\G}[1]{\Gamma_{#1}}

\newcommand{\X}[1]{\hat{X}_{#1}}
\newcommand{\Y}[1]{\hat{Y}_{#1}}
\newcommand{\Xp}[1]{\dot{\hat{X}}_{#1}}
\newcommand{\Yp}[1]{\dot{\hat{Y}}_{#1}}

\newcommand{\q}{\hat{q}}
\newcommand{\susc}[1]{\chi_\mathrm{#1}}
\newcommand{\etam}{\eta_\mathrm{meas}}

\newcommand{\Sqq}{\overline{S}_{\hat{q}\hat{q}}}

\newcommand{\nocontentsline}[3]{}
\newcommand{\tocless}[2]{\bgroup\let\addcontentsline=\nocontentsline#1{#2}\egroup}

\begin{document}


\title{\bf Entanglement of Propagating Optical Modes via a Mechanical Interface}

\author{Junxin Chen,$^{1,2,\ast}$ Massimiliano Rossi,$^{1,2,\ast}$ David Mason,$^{1,2,3,\ast}$ Albert Schliesser,$^{1,2,\dagger}$\\
\\
\footnotesize{$^{1}$Niels Bohr institute, University of Copenhagen, 2100 Copenhagen, Denmark}\\
 \footnotesize{$^{2}$Center for Hybrid Quantum Networks (Hy-Q), Niels Bohr Institute, University of Copenhagen,}\\
 \footnotesize{2100 Copenhagen, Denmark}\\
 \footnotesize{$^{3}$Present address: Department of Applied Physics, Yale University, New Haven, CT, USA}\\
\\
\footnotesize{$^\ast$these authors contributed equally to this work}\\
\footnotesize{$^\dagger$To whom correspondence should be addressed; E-mail:  albert.schliesser@nbi.ku.dk}
}

\date{}
\baselineskip18pt
\maketitle
\begin{abstract}
Many applications of quantum information processing (QIP) require distribution of quantum states in networks, both within and between distant nodes\cite{kimble2008quantum}. 
Optical quantum states are uniquely suited for this purpose, as they propagate with ultralow attenuation and are resilient to ubiquitous thermal noise. 
Mechanical systems are then envisioned as versatile interfaces between photons and a variety of solid-state QIP platforms\cite{stannigel2010optomechanical,Kurizki2015}.
Here, we demonstrate a key step towards this vision, and generate entanglement between two propagating optical modes, by coupling them to the same, cryogenic mechanical system. 
The entanglement persists at room temperature, where we verify the inseparability of the bipartite state and fully characterize its logarithmic negativity by homodyne tomography. 
We detect, without any corrections, correlations corresponding to a logarithmic negativity of $E_\mathrm{N}=0.35$.
Combined with quantum interfaces between mechanical systems and solid-state qubit processors already available\cite{reed2017faithful,pirkkalainen2013hybrid,satzinger2018quantum,chu2018creation} or under development\cite{rabl2010quantum,kolkowitz2012coherent}, this paves the way for mechanical systems enabling long-distance quantum information networking over optical fiber networks.

\end{abstract}

Entanglement is a crucial resource for QIP\cite{horodecki2009quantum}. 
As such, the ability to entangle fields of arbitrary wavelength will be important for linking nodes in heterogeneous QIP networks.  
Mechanical oscillators are uniquely poised in their ability to create such links, thanks to the frequency-independence of the radiation pressure interaction. 
%
%
The ability to entangle two radiation fields via a common mechanical interaction
was outlined 20 years ago\cite{giovannetti2001radiation, giannini2003}, and the intervening decades have seen the development of optomechanical devices\cite{aspelmeyer2014cavity} which are robustly quantum mechanical and routinely integrated into hybrid systems. 

%
Recently, mechanically-mediated entanglement has been reported between propagating  microwave \linebreak fields\cite{barzanjeh2019stationary} as well as two superconducting qubits\cite{bienfait2019}.
In both cases, the entanglement remained confined to the dilution refrigerator in which it was created.  
Here, we utilize an extremely coherent mechanical platform to create, for the first time, mechanically-entangled \textit{optical} fields, spanning up to $100\,$nm in wavelength. 
Moreoever, while the entangling mechanical interface resides at cryogenic temperatures, it is compatible with highly-efficient light extraction and collection, such that we can directly measure the entanglement at room temperature, without noise subtraction or other indirect inference.  
This in turn means that the entangled optical fields could easily be distributed for QIP applications.  

We consider two propagating optical fields (labelled by $j=A,B$), from which one can identify a pair of temporal modes with quadratures $\hat{X}_j,\hat{Y}_j$.  
We take the variance of these modes to be 1/2 in their ground state.  
From these modes, one can construct joint EPR-type variables $\hat{X}_\pm=\hat{X}_A\pm\hat{X}_B$ and $\hat{Y}_\pm=\hat{Y}_A\pm\hat{Y}_B$, which form the basis for various entanglement criteria\cite{duan2000inseparability,giovannetti2003characterizing}.  
We adopt the common DGCZ criterion\cite{duan2000inseparability} for the inseparability $\mathcal{I}$, which states that the two modes are inseparable if their variances ($V$) satisfy:
\begin{eqnarray}
    \mathcal{I}\equiv\frac{V(\hat{X}_+)+V(\hat{Y}_-)}{2}<1.
\end{eqnarray}

To further quantify this entanglement, one can utilize the system's covariance matrix, $\boldsymbol{\sigma}$, which fully characterizes the correlations between various quadratures.
From this matrix, it is straightforward to calculate the sympectic eigenvalues of its partial transpose, $\tilde{\nu}_\pm$\cite{adesso2004extremal}.  
These eigenvalues offer a condition for separability ($2\tilde{\nu}_- \ge 1$), as well as a tool to calculate a common measure of entanglement, the logarithmic negativity $E_N=\mathrm{max}\left[0,-\mathrm{log_2}\,2\tilde{\nu}_-\right]$.  
We note that 2$\tilde{\nu}_-$ also corresponds to the minimum value of $\mathcal{I}$ possible when optimizing over local operations on either subsystem (e.g. squeezing, rotation)\cite{zippilli2015entanglement}.  
Thus, 2$\tilde{\nu}_-$ serves as a lower bound for any DGCZ measurement. 

\begin{figure}[!t]
\begin{center}
\includegraphics[width=88mm]{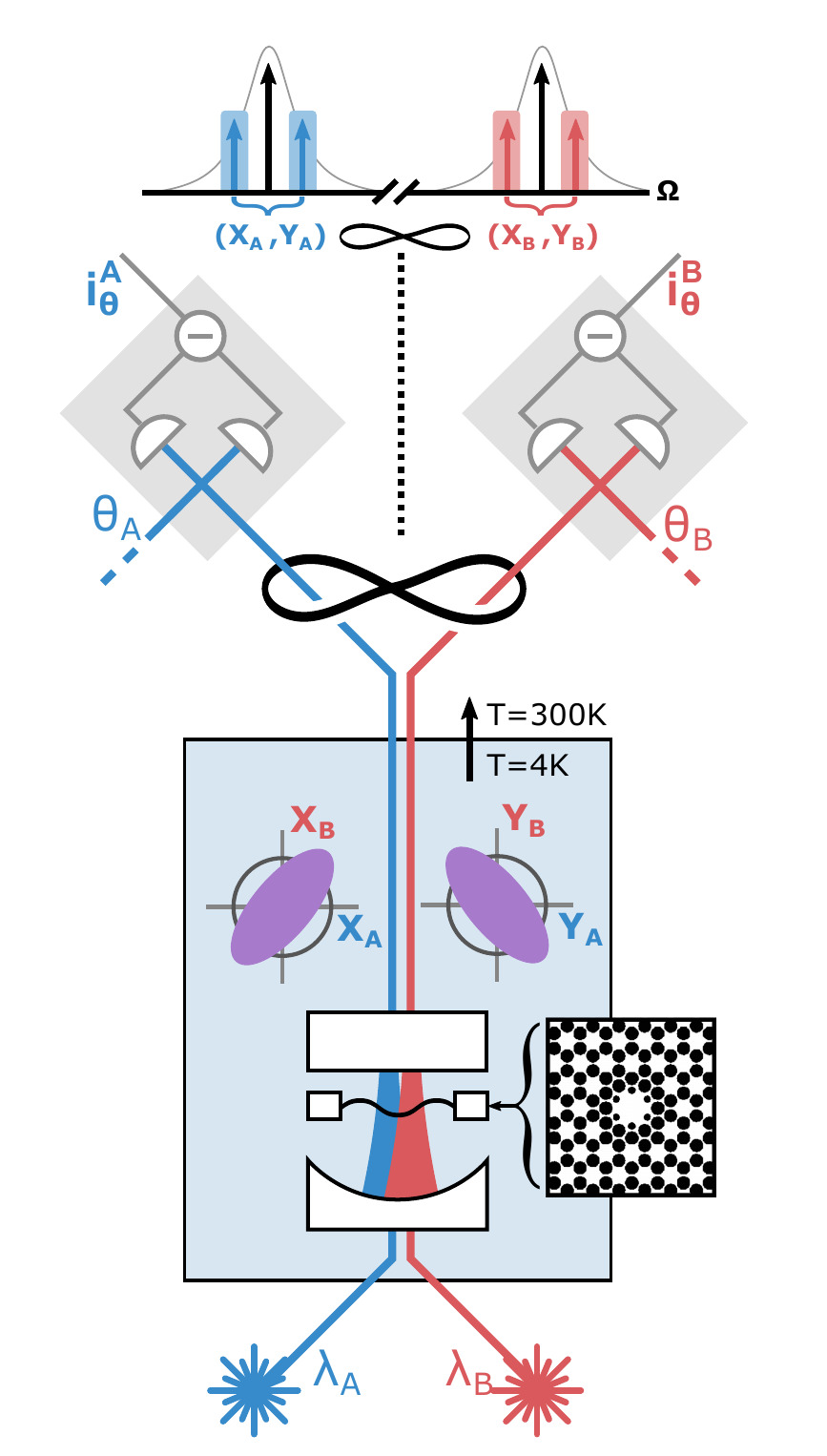}
\caption{{\bf Experimental Setup} Two lasers (red, blue) simultaneously drive an optomechanical cavity, kept in a helium flow cryostat. The inset shows the structure of the soft-clamped mechanical resonator ($\mathrm{Si}_3\mathrm{N}_4$ in white, holes in black). Exiting the cavity, the optical fields possess non-local correlations, illustrated by the squeezed phase space ellipses. After the cavity, the two lasers are physically separated and detected simultaneously by balanced homodyne detectors, with local oscillators locked at phases $\theta_{A},\theta_{B}$.  The top of the figure shows a frequency diagram of the relevant optical modes.  The two cavity drives are shown in black, with scattered mechanical sidebands of laser $A$ and $B$ shown in blue and red, respectively.  The sideband quadrature modes considered in the manuscript correspond to combinations of both scattered sidebands, as indicated by the blue and red shaded areas.}
\label{f:setup}
\end{center}
\end{figure}

In an optomechanical setting, in which a mechanical resonator (unitless position $\hat{q}$, momentum $\hat{p}$) is linearly coupled to two independent, resonantly-driven cavity modes (amplitude $\hat{X}^{\mathrm{cav}}_j$, phase $\hat{Y}^{\mathrm{cav}}_j$, $j=A,B$), 
the interaction Hamiltonian can be written as:
$\hat{H}_\mathrm{int}=-\sum_{j}2\hbar g_j \hat{X}^\mathrm{cav}_j\hat{q}$,
where $\hbar$ is the reduced Planck constant, and $g_j$ are the field-enhanced optomechanical coupling rates. 
We consider an unresolved-sideband system, in which the cavity decay rates, $\kappa_j$ are much faster than the mechanical frequency $\Om$ and mechanical energy dissipation rate, $\Gm$.  
We also assume that the two cavity modes are driven symmetrically by coherent states, such that their induced quantum backaction rates, $\Gamma_j^\mathrm{qba}=4g_j^2/\kappa_j$ are equal: $\Gamma_A^\mathrm{qba}=\Gamma_B^\mathrm{qba}\equiv\Gamma_\mathrm{qba}$. 
The quadratic interaction preserves the Gaussianity of the state.
The following equations of motion link the input and output optical fields:
\begin{linenomath*}
\begin{subequations}\label{e:EOM}
\begin{align}
\hat{X}_j^\mathrm{out}&(t) = -\hat{X}_j^\mathrm{in}(t)\\
\hat{Y}_j^\mathrm{out}&(t) = -\hat{Y}_j^\mathrm{in}(t) - 2\sqrt{\Gamma_\mathrm{qba}}\chi_\mathrm{m}(t)*[\sqrt{2\G{m}}\hat{P}_\mathrm{in}(t) + \sum_{i=A,B}2\sqrt{\Gamma_\mathrm{qba}}\hat{X}_i^\mathrm{in}(t)],
\end{align}
\end{subequations}
\end{linenomath*}
where $\hat{X}_j^\mathrm{in}$ and $\hat{Y}_j^\mathrm{in}$ are input vacuum noise quadratures, $\chim$ is the mechanical susceptibility, $\G{\mathrm{m}}$ is the mechanical energy dissipation rate, $*$ indicates convolution, and $\hat{P}_\mathrm{in}$ is the mechanical thermal noise operator.

In equations (\ref{e:EOM}), we see that the quantum amplitude fluctuations of each laser drive the mechanical system, whose motion is then imprinted on the optical phase. 
This is the same mechanism that drives ponderomotive squeezing of a single laser\cite{safavi2013squeezed}, but in this case there are also cross-correlations between the lasers.  
More insight can be had by moving to the joint mode basis (see Supplementary), where one finds that the system decouples into a sum mode undergoing ponderomotive squeezing, and a difference mode which remains dark to all mechanical dynamics. 
It is this squeezing of a joint (non-local) mode which results in the ``ponderomotive entanglement'' we study here. 
We note that equations~(\ref{e:EOM}) also closely mirror those describing four-mode squeezing based on the Kerr nonlinearity in glass\cite{levenson1987four}, in which the response function $\chim$ is effectively instantaneous.
Both approaches are members of a broader class of settings that enable, in principle, quantum-non-demolition measurements of light\cite{grangier1998quantum,pontin2018quantum}.

Homodyne detection allows measurement of optical quadratures in a rotated basis defined by the local oscillator phase, $\theta_j$.  
Filtering the homodyne signal at frequency $\Omega$ with a mode function $h(t)$ yields the sideband quadratures of a particular temporal mode at time $t$:
\begin{linenomath*}
\begin{subequations}
\begin{align}
    \hat{X}_{j}^{\theta_j}(t)&=\int^t_{-\infty}ds\, \mathrm{cos}(\Og s)h(t-s)\left(\hat{X}_j^\mathrm{out}(s)\cos(\theta_j)+\hat{Y}_j^\mathrm{out}(s)\sin(\theta_j)\right)\\
    \hat{Y}_{j}^{\theta_j}(t)&=\hat X_j^{\theta_j+\pi/2}(t),
\end{align}
\end{subequations}
\end{linenomath*}
where $\theta_j$ is the homodyne angle.
Note that the quadratures $\hat{X}_{j}^{\theta_j}$ available in a homodyne
detector contain a \emph{pair} of sidebands, symmetric to the
carrier\footnote{Thus correlations between such modes are sometimes called
  four-mode-squeezing.} (see Supplementary), in contrast to the entangled microwave modes recently analyzed in a heterodyne scheme\cite{barzanjeh2019stationary}.
In the following model, we consider the limit of long filter times, in which $h$ effectively selects a single Fourier component\cite{zippilli2015entanglement}.
Furthermore, since the system is stationary, we drop the time argument $t$ and focus on the ensemble statistics of these modes.

Within this model, one can obtain a simple expression for the DGCZ inseparability criterion,
\begin{eqnarray}\label{e:duanToy}
\mathcal{I}^\mathrm{ideal}_{\Theta,\Omega} = 1 
+ 8\Gamma_\mathrm{qba}|\chim(\Og)|^2 \G{\mathrm{dec}}\left(1 + \cos(2\Theta)\right) 
- 4\Gamma_\mathrm{qba}\mathrm{Re}\left[\chim(\Og)\right]\sin(2\Theta),
\end{eqnarray}
where $\Theta \equiv(\theta_A+\theta_B)/2$.
The first term is the contribution from shot noise at the detectors. The second term is the contribution from mechanical motion, where the total decoherence rate $\G{\mathrm{dec}}=2\Gamma_\mathrm{qba}+\G{m}(\bnth+1/2)$ includes both quantum backaction sources and thermal motion. 
The third term corresponds to correlation between two beams, again in close analogy to ponderomotive squeezing\cite{safavi2013squeezed}.
In practice, there is always optical loss, which admits  vacuum noise that degrades the detected correlations.
This is described by a collection efficiency $\eta_A=\eta_B\equiv\eta<1$, with which the inseparability of the detected optical states becomes $\mathcal{I}_{\Theta,\Omega}=\eta \mathcal{I}^\mathrm{ideal}_{\Theta,\Omega}+(1-\eta)$.
By defining a combined measurement efficiency $\etam=2\eta\Gamma_\mathrm{qba}/\G{\mathrm{dec}}$, one can show (see Supplementary) that the minimum of $\mathcal{I}_{\Theta,\Omega}$ is given by $1-\etam/2$. 
By further calculating the full covariance matrix for this toy model (see Supplementary), one can show that $\min\{2\tilde{\nu}_-\}=\sqrt{1-\etam}$, that is, the system can generate arbitrarily strong entanglement as $\etam\rightarrow1$.
%
%
%
%
%
%
%
%
%
%
%
%
%

In practice, the optical fields become entangled via their shared interaction with a $3.6\,\mathrm{mm}\times3.6\,\mathrm{mm}\times20\,\mathrm{nm}$ soft-clamped 
$\mathrm{Si_3N_4}$ membrane\cite{tsaturyan2017ultracoherent}. 
The vibrational mode of central defect has a frequency of $\Om=2\pi\times1.139\,\mathrm{MHz}$, and a quality factor $Q=1.04\times10^9$ at a temperature of $10\,\mathrm{K}$, which corresponds to a mechanical linewidth of $\Gm=2\pi\times1.10\,\mathrm{mHz}$.

The membrane is inserted in the middle of an optical cavity\cite{thompson2008aa, rossi2018measurement}, addressed by two lasers with wavelength $\sim796$~nm. 
These lasers are orthogonally polarized and populate the cavity in two different longitudinal modes separated by $\sim0.3$~THz, with linewidths of $\kappa_A = 2\pi\times13.3$~MHz and $\kappa_B = 2\pi\times12.6$~MHz.
With this setup, we achieve $\Gamma^\mathrm{qba}_A\approx2\pi\times1.35$~kHz and $\Gamma^\mathrm{qba}_B\approx2\pi\times0.89$~kHz, which easily exceed the thermal decoherence rate $\Gm\bnth\approx2\pi\times0.20$~kHz.
We measure the optical quadratures of the cavity output fields using two separate balanced homodyne detectors, achieving overall collection efficiencies of $\eta_A = 60\%$ and  $\eta_B = 77\%$. 
This gives a combined measurement efficiency of $\etam=58\%$ (see Supplementary).

By combining slope and dither lock techniques we are able to arbitrarily stabilize the phase of the local oscillators in the range $[0, 2\pi)$. 
The photocurrent of each balanced homodyne detector is digitized with a 15 MSa/s ADC. 
We numerically demodulate the photocurrents at frequency $\Og/(2\pi)=1.1416\,\mathrm{MHz}$, and low-pass filter the result with bandwidth $200\,\mathrm{Hz}$. 
(This bandwidth is narrow compared to the mechanical feature of interest, allowing us to apply the infinitely-long-filter limit of the model.)

\begin{figure}[!t]
\begin{center}
\includegraphics[width=0.99\columnwidth]{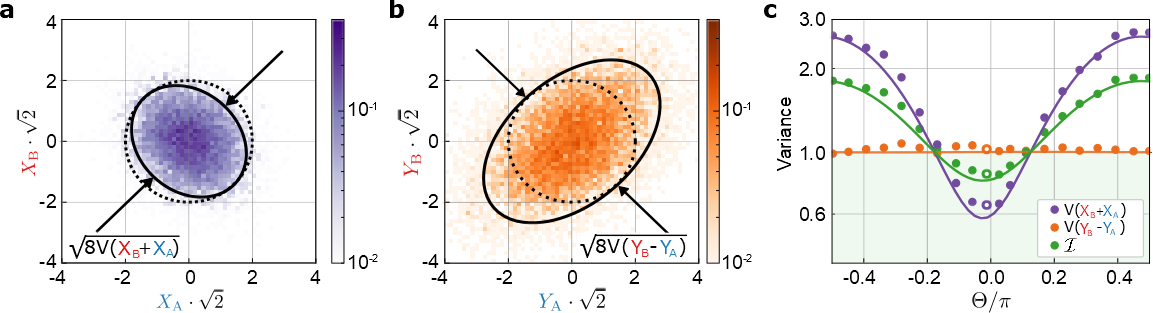}
\caption{{\bf EPR quadrature statistics.} {\bf a},
  {\bf b}, 2D histograms of raw $X$ and $Y$ quadrature data, respectively, for
  $\Theta\approx0$. The black dashed circle indicates (2$\times$) the standard deviation of vacuum
  noise, which has a radius of $1/\sqrt{2}$ (note the axes' scale factor of $\sqrt{2}$). The solid black ellipses indicate (2$\times$) the covariance ellipse of the measured data.  The black arrows indicate the diagonal/anti-diagonal standard deviations which are relevant for calculating the DGCZ criterion.
{\bf c}, Homodyne angle dependence of joint quadrature variances. The purple
(orange) dots are the sum (difference) quadrature $\hat{X}_+$ ($\hat{Y}_-$). The
average of these yields the DGCZ inseparability criterion (green points).  The measurement ensembles contain $\sim10^4$ samples, such that the statistical
standard error of the variance estimators is $\sim$2\% of the reported values. This contains both the error in the estimation of the EPR variances and the error in the estimations of the vacuum noise variance. The solid line is the theoretical prediction.}
\label{f:entanglement}
\end{center}
\end{figure}

We now proceed to characterize the variance of EPR-type variables, as introduced above, to compare with the DGCZ criterion.
We choose a common basis $\theta_{A}\approx\theta_{B}\approx\Theta\approx 0 $ and measure, in sequence, the combinations $\{\hat{X}_A^\Theta,\hat{X}_B^\Theta\}$, $\{\hat{Y}_A^\Theta,\hat{Y}_B^\Theta\}$, and vacuum noise (by blocking the cavity output).
Figures \ref{f:entanglement} {\bf a.} and {\bf b} show histograms of the measured quadrature data for the $X$ and $Y$ quadratures, along with reference lines for vacuum noise variance in black. 
Recalling the joint quadrature definitions, we note that the DGCZ criterion involves the diagonal and anti-diagonal variances of the $X$ and $Y$ histograms, respectively.  
In the figure, we clearly see squeezing in the former, and near-vacuum variance in the latter -- thus already indicating violation of the DGCZ criterion. 
Quantitatively, we find
$\mathcal{I}=0.83\pm2\%\mathrm{(stat.)}\pm0.3\%\mathrm{(syst.)}$. 
The statistical error comes from the number of samples used to estimate the EPR variances and the vacuum noise. The systematic error arises from the estimations of the vacuum noise variance, due to residual classical amplitude noise and mismatch in the photodiode responsivities, at the balanced homodyne detectors (see Supplementary).
We also notice that the variances in the orthogonal directions are
at the vacuum  level. This does not violate the Heisenberg
uncertainty relation, since the pairs of quadratures \{$\hat{X}_A, \hat{X}_B$\}
and \{$\hat{Y}_A, \hat{Y}_B$\} commute with each other and are not canonically conjugate
observables.
We repeat such measurement for different homodyne angles ($\Theta \in [-\pi/2,\pi/2]$) as shown in Figure \ref{f:entanglement} {\bf c}. 
The solid lines are theoretical predictions based on a full model of optomechanical dynamics (taking, in particular, dynamical backaction\cite{aspelmeyer2014cavity} into account), using system parameters extracted from fits (see Appendix).
We find good agreement over all phases, firmly establishing that the two optical modes satisfy the DGCZ inseparability criterion.
From Figure~\ref{f:entanglement} {\bf c} we also notice that the best two-mode squeezing we achieve is, for the quadrature $\hat{X}_+$, 1.8 dB below the vacuum noise limit.



\begin{figure*}[!tt]
\begin{center}
\includegraphics[width=1\columnwidth]{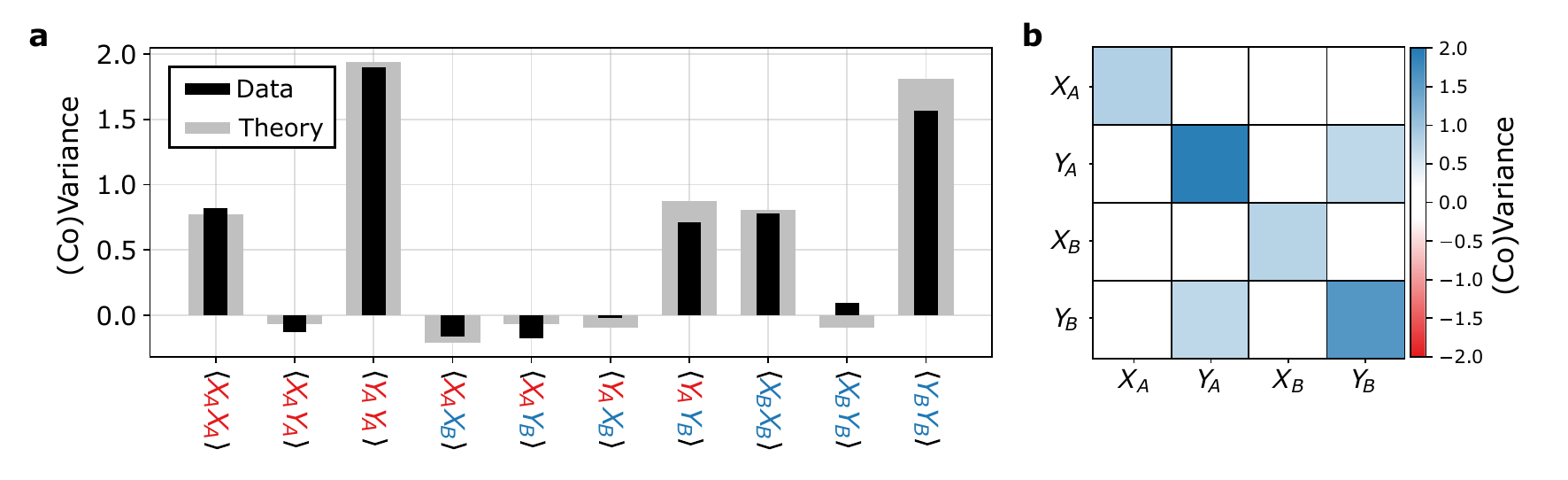}
\caption{{\bf Covariance matrix of the two optical modes}. {\bf a}, Measured (black) and predicted (gray) entries of the covariance matrix. The measurement ensembles contain $\sim10^4$ samples, such that the standard error of the variance estimators is $\sim$2\% of the reported values. {\bf b}, Matrix-representation of the measured data, to highlight the location of the significant non-zero entries.}
\label{f:CovMat}
\end{center}
\end{figure*}

Having established entanglement, we now quantify it by reconstructing the covariance matrix by Gaussian homodyne tomography.  
By measuring 5 different pairs of angles $\{\theta_{A},\theta_{B}\} = \{0,0\},\{\pi/2,\pi/2\}$,\quad$\{0,\pi/2\},\{\pi/2,0\},\{\pi/4,\pi/4\}$, we obtain all necessary intra-system and inter-system correlations. 
The reconstructed covariance matrix and theoretical prediction are shown in Figure~\ref{f:CovMat}. 
From this experimental data, we find a minimum symplectic eigenvalue $2\tilde{\nu}_-=0.79$ , corresponding to $E_\mathrm{N}=0.35$.

\begin{figure*}[!ht]
\begin{center}
\includegraphics[scale=1]{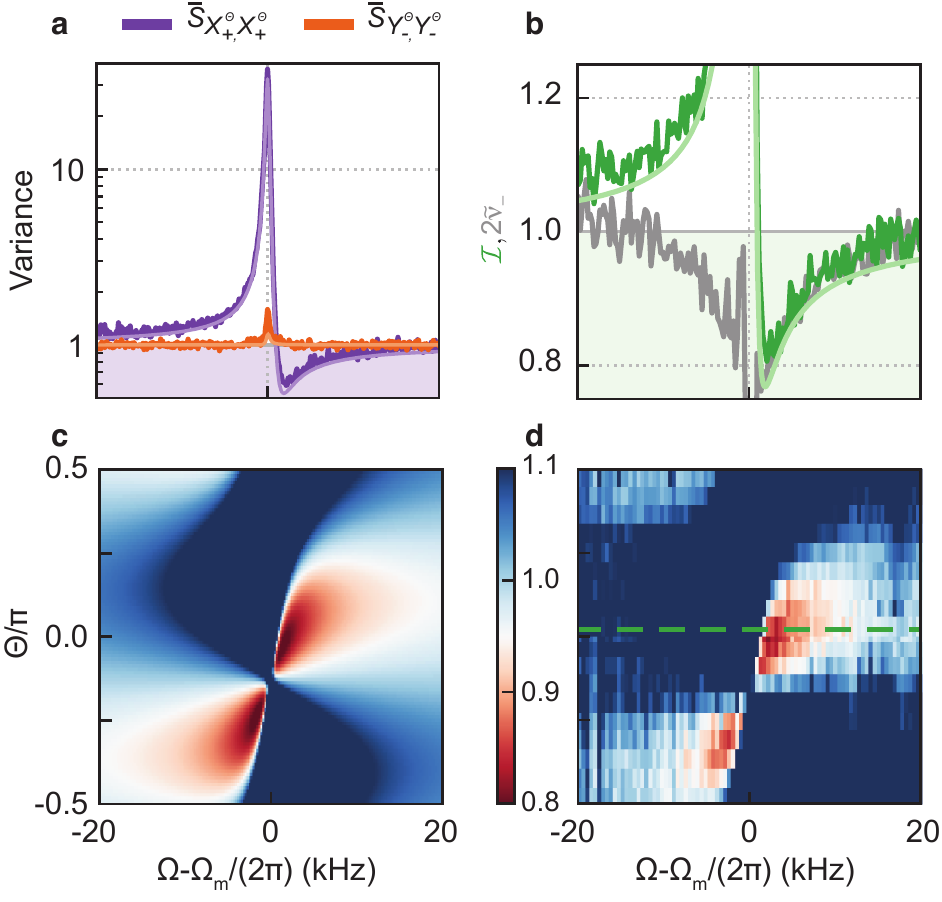}
\caption{{\bf Frequency-dependent entanglement}.
{\bf a}, Variance of the EPR-type joint quadratures, $\hat{X}_{+}^\Theta$ (purple) and $\hat{Y}_{-}^\Theta$ (orange), from sum and difference of the two measured homodyne photocurrents, at angle $\Theta\approx0$ and {\bf b} derived inseparability (green), as a function of center frequency $\Omega$. The solid lines in {\bf a} and {\bf b} are fit to a full model (see Supplementary). The dark gray line represents the minimum symplectic eigenvalue $2\tilde{\nu}_-$, obtained by optical homodyne tomography.
The modes of the two laser fields are entangled whenever $2\tilde{\nu}_-<1$.
{\bf c}, Theory and {\bf d}, measurements of the inseparability $\mathcal{I}(\Theta, \Omega)$. The green dashed line indicates the measurement shown in {\bf b}.
The horizontal axes are referenced to the bare mechanical frequency $\Om=2\pi\times1.139$~MHz.
}
\label{f:espec}
\end{center}
\end{figure*}

The previous results refer to a sideband quadrature mode at a particular frequency, $\Og$. 
We now examine how this entanglement varies as we sweep $\Og$ near the mechanical resonance, $\Om$.
(Note that for computational convenience, we do this by calculating noise spectral densities via the Fast Fourier Transform (FFT) of the raw photocurrents, which corresponds to a mode function $h(t)=\theta(t)\theta(T-t)/\sqrt{T}$ where $\theta(t)$ is the Heaviside function and $T\approx9$~ms is the acquisition time). 
Figure~\ref{f:espec} shows such a frequency-dependent inseparability, as well as its dependence on the homodyne measurement basis, $\Theta$.

We see that the entanglement criteria can be met for frequencies above and below mechanical resonance, in a manner consistent with the dispersive third term in equation~(\ref{e:duanToy}).
The solid lines in Figure~\ref{f:espec}{\bf a} and {\bf b} are theoretical predictions from the full model, based on a single set of system parameters, obtained from independent measurements or fits to account for drifts (see Appendix).
Moreover, similar to the measurement in Figure~\ref{f:CovMat}, we can reconstruct the covariance matrix (and corresponding $\tilde{\nu}_-$) for each frequency bin, as shown in Figure \ref{f:espec}{\bf b}.
We see that, as expected, $2\tilde{\nu}_-$  serves as a lower bound for the inseparability $\mathcal{I}$. 
Since this bipartite Gaussian state is approximately symmetric, from $2\tilde{\nu}_-$ we can calculate the entanglement of formation, which is accepted as a proper measure of quantum correlations as a resource\cite{barzanjeh2019stationary, flurin2012aa,tserkis2017quantifying}.
Integrating this quantity over a 30 kHz bandwidth yields an entanglement distribution rate of 753~ebits/s.


\begin{figure*}[!ht]
\begin{center}
\includegraphics[scale=1]{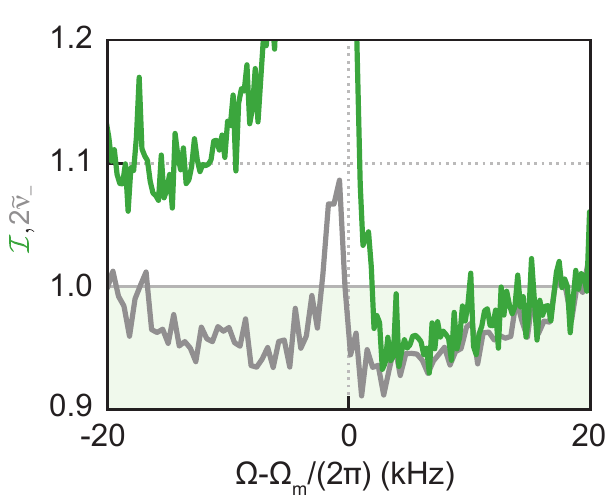}
\caption{{\bf Entanglement spanning 120~nm in wavelength.}
Inseparability $\mathcal{I}(\Theta\approx0, \Omega)$ (green) and minimum of the symplectic eigenvalue $2\tilde{\nu}_-$ (gray) from homodyne tomography. The horizontal axis is referenced to the bare mechanical frequency $\Om=2\pi\times1.139$~MHz.
}
\label{f:800-900}
\end{center}
\end{figure*}

We emphasize that the optomechanical interaction which generated the entanglement presented above is fundamentally wavelength-independent.
To illustrate this, we move laser A to $\sim912$~nm, and repeat the measurements of Figure~\ref{f:espec}. 
We observe a DGCZ variance below unity and a minimum symplectic eigenvalue $2\tilde{\nu}_-=0.92$ for a mode centered at $\Og=2\pi\times1.142$~MHz with bandwidth $915$~Hz. 
The performance is degraded compared to the previous results, due to less efficient light collection at $\sim912$~nm. 
Nevertheless, these results establish entanglement of two lasers separated by more than $100\,\mathrm{nm}$ in wavelength. 
%
%
%
%
%
%
%
%
%
%
%
%
%
%

In conclusion, we have demonstrated quadrature entanglement of two non-degenerate optical beams via their common radiation-pressure interaction with a mechanical resonator.
While applications in optical quantum communication are conceivable (as realized with entangled light sources based on more traditional optical parametric oscillators \cite{ou1992realization,Silberhorn2001,villar2005generation,grosse2008observation}), mechanical platforms offer unique possibilities.
In particular, the combination of mechanically mediated microwave\cite{barzanjeh2019stationary} and optical (this work) entanglement could enable microwave-optical entanglement, based on membrane electro-opto-mechanical systems \cite{bagci_optical_2014, higginbotham2018}. 
This would constitute a much-needed resource for networks of quantum computers based on superconducting qubits. 
In this context, it is noteworthy that the mechanical interface can in principle also store quantum information. Indeed, for the device presented here, the memory time is ca. 1 ms even for 10 K operation\cite{rossi2018measurement}, easily exceeding storage time in optical fibers. 

In our work, entanglement is preserved from the cryogenic mechanical mediator all the way to the laser beams analyzed in room-temperature homodyne detectors. 
This enhances the prospects of a general class of hybrid quantum systems\cite{Kurizki2015, stannigel2010optomechanical} based on mechanical interfaces, which could harness entanglement between solid-state (e.g. spin or charge-based) quantum systems, typically operating at low temperatures, and itinerant optical fields. 

From a more fundamental perspective, it would be interesting to explore concepts at the interface of quantum measurement and entanglement. 
For instance, the optomechanical interaction in this work can also be interpreted as a strong quantum measurement of the mechanics. 
This system should be well-suited for studying the usually-hard-to-access system-meter entanglement\cite{vitali2007aa, krisnanda2017revealing}.


\clearpage
{\bf Acknowledgments} The authors acknowledge sample fabrication by Y. Tsaturyan and fruitful discussions with E.~Zeuthen, A.~S.~S{\o}rensen and J.~Appel. This work was supported by funding from the European Union’s Horizon 2020 research and innovation programme (European Research Council project Q-CEOM, grant agreement no. 638765 and FET proactive project HOT, grant agreement no. 732894).

{\bf Author Contributions} J.~C., M.~R. and D.~M. built the set-up and performed the experiments, analysed the data and, together with A.S., discussed the results and wrote the paper. A.S. supervised the project.

{\bf Competing Interest} The authors declare no competing interests.

{\bf Data availability} The data that support the findings of this study are available from the corresponding author upon reasonable request.

{\bf Code availability} The code used in the analysis of the data is available from the corresponding author upon reasonable request.


\clearpage
\newpage
\bibliography{bibliography}

\renewcommand{\figurename}{{\bf Supplementary Fig.}}
\renewcommand{\tablename}{{\bf Supplementary Table}}
\setcounter{figure}{0}\renewcommand{\thefigure}{{\bf S\arabic{figure}}}
\setcounter{table}{0}\renewcommand{\thetable}{{\bf S\arabic{table}}}
\setcounter{equation}{0}\renewcommand{\theequation}{S\arabic{equation}}

\baselineskip20pt

%
%
%
%
%
%
%
%
\clearpage
\newpage

{\Huge Supplementary Information}

\tableofcontents

\clearpage
\newpage
\section{Theory}
Here we present a theoretical analysis of the experiments described in the main text.
First, we derive a model which fully describes the dynamics of an optomechanical system composed of two optical modes and a single mechanical mode. 
From these equations we derive a simplified, toy model, useful for the understanding of the physics. 
For the reader interested only in the toy model, we point out that it can be obtained from the complete equations in the limit of resonant lasers ($\Delta_j=0$), unresolved cavities ($\kk{j}\gg\Om$) and identical measurement rates ($\Gamma^\mathrm{meas}_A=\Gamma^\mathrm{meas}_B=\Gmeas$). 
Then we move to describe how one can fully characterize the entanglement of the state from the covariance matrix. 
Next, we interpret the optical-optical entanglement as joint quadrature squeezing, in the framework of toy model. 
Finally, we illustrate the relation between the entangled optical modes and the ones measured by balanced homodyne detectors. 
\\
\subsection{Three-mode optomechanical system}
We consider two cavity modes $A$ and $B$ at optical frequency $\omega_{\mathrm{c}_{j}}$, described by the ladder operator $\hat{A}_j^\mathrm{cav} = \alpha_j^\mathrm{cav} + \hat{a}_j^\mathrm{cav}$ ($j=A,B$), where $\hat{a}_j^\mathrm{cav}$ is an operator representing fluctuations around the coherent amplitude $\alpha_j^\mathrm{cav}$ and satisfying canonical commutation relations $[\an{j}(t), \andag{i}(t') ] = \delta_{ji} \delta(t-t')$. 
In the following, we assume that the cavity field $\alpha_j^\mathrm{cav}$ is real. 
The mechanical resonator mode is described by the dimensionless position $\hat{Q} = Q + \hat{q}$ and its momentum $\hat{p}$. 
The fluctuations around the mean displacement $Q$ are described by the operator $\hat{q}$, which also satisfies canonical commutation relations $[\q(t), \hat{p}(t') ] = i \delta(t-t')$.\\
The optomechanical coupling is described by the rates $\vcrn{j}$. In order to effectively enhance such coupling rate, we displace the cavity field by means of coherent driving lasers $\hat{A}_j^\mathrm{in,L} = (\alpha_j^\mathrm{in} +\hat{a}_j^\mathrm{in,L}) e^{-i(\omega_{\mathrm{L}, j}t+ \phi_j)}$. 
The phase $\phi_j$ is needed in order to be consistent with the convention adopted of real intracavity field. 
We move each cavity modes to a frame rotating at the lasers' frequency $\omega_{\mathrm{L}, j}$. 
In the limit of large coherent optical field $\alpha_j$, the interaction can be linearized and the Hamiltonian in this interaction picture reads
\begin{eqnarray}\label{e:lin_H}
    \hat{H}_\mathrm{IP} = \hbar\Om \frac{\hat{q}^2+\hat{p}^2}{2}
    -\sum_{j}\hbar\Delta_j\andag{j}\an{j}-\sum_{j}\sqrt{2}\hbar g_j(\andag{j}+\an{j})\hat{q},
\end{eqnarray}
where $\Delta_j = \tilde{\Delta}_j+\sqrt{2} \vcrn{j} Q = \omega_{\mathrm{L}, j} - \omega_{\mathrm{c}_{j}}$ is the detuning between the laser field and the cavity modes and $g_j = \vcrn{j} \alpha_j^\mathrm{cav}$ is the enhanced optomechanical coupling. 
\\
The cavity modes are coupled to a lossy environment through two ports, hereby indicated as $L$ and $R$, at rates $\kk{j,L}$ and $\kk{j,R}$ respectively. The total damping rate is $\kk{j} = \kk{j,L} + \kk{j,R}$.
The mechanical system is also  coupled to an environment, which leads to a damping rate $\Gm$. 
We describe these open dynamics via the quantum Langevin equations. 
We also move to the quadrature representations of the optical modes, defined as $\hat{X}_j = (\hat{a}_j^\dagger + \hat{a}_j)/\sqrt{2}$ and $\hat{Y}_j = i ( \hat{a}_j^\dagger - \hat{a}_j )/\sqrt{2}$. The equations of motions are
\begin{linenomath*}
\begin{subequations}\label{e:qle_time}
\begin{align}
\Xp{j} = &-\frac{\kk{j}}{2} \X{j} - \Detn{j} \Y{j}  +\sqrt{\kk{j,L}}  \hat{X}^\mathrm{in,L}_{j,\phi_j}+  \sqrt{\kk{j,R}} \hat{X}^\mathrm{in,R}_j,\\
\Yp{j} = &-\frac{\kk{j}}{2} \Y{j} + \Detn{j} \X{j} +2 g_{j} \hat{q} + \sqrt{\kk{j,L}} \hat{Y}^\mathrm{in,L}_{j,\phi_j}+  \sqrt{\kk{j,R}} \hat{Y}^\mathrm{in,R}_j,\\
\ddot{\q} = & - \G{m}\dot{\q} - \Om^2\q + \Om\left(\sum_j 2g_{j} \X{j} + \sqrt{2\G{m}}\Fth\right).
\end{align}
\end{subequations}
\end{linenomath*}
The cavity modes are driven by optical vacuum noise $(\hat{X}^\mathrm{in,L}_{j,\phi_j}, \hat{Y}^\mathrm{in,L}_{j,\phi_j})$ and $(\hat{X}^\mathrm{in,R}_j, \hat{Y}^\mathrm{in,R}_j)$, where $\phi_j = \tan^{-1}(\frac{\Delta_j}{\kk{j}/2})$ is the rotation due to the cavity response.
The only non-zero symmetrized correlations are of the form $\overline{\langle\hat{X}_i(t) \hat{X}_j(t')\rangle}=\overline{\langle\hat{Y}_i(t) \hat{Y}_j(t')\rangle}=\frac{1}{2}\delta_{ij} \delta(t-t')$\footnote{An overlined quantity refers to its symmetrized version, i.e. $\overline{\langle\hat{X}_i(t) \hat{X}_j(t')\rangle}\equiv\left(\langle\hat{X}_j(t) \hat{X}_i(t')\rangle+\langle\hat{X}_i(t') \hat{X}_j(t)\rangle\right)/2$.}. 
The mechanical system is driven by a Brownian force $\Fth$. In the regime of underdamped motion $(\Gm\ll\Om)$ and high-temperature of the bath $(\kB T\gg\hbar\Om)$\cite{giovannetti2001aa}, this noise becomes Markovian and its symmetrized correlations is $\overline{\langle\Fth(t)\Fth(t')\rangle}=\left(\bnth+1/2\right)\delta(t-t')$.
\\
We can experimentally only measure a field which leaks out from the cavity. 
We describe it with input-output relations, according to which the output field of mode $j$ through the port $i$ is 
\begin{eqnarray}\label{e:in-out}
\hat{A}_j^{\mathrm{out}, i} = \hat{A}_j^{\mathrm{in}, i} - \sqrt{\kk{j,i}}\hat{A}_j^\mathrm{cav}.
\end{eqnarray}
The systems of Equation~\eqref{e:qle_time} and \eqref{e:in-out} can be promptly solved in the Fourier domain\footnote{For an operator $\hat{a}(t)$, we define its Fourier transform $\hat{a}(\Og)\equiv\int_{-\infty}^{\infty}dt \mathrm{e}^{i \Og t} \hat{a}(t) $.}. 
Let's introduce the cavity quadratures' susceptibilities
\begin{linenomath*}
\begin{subequations}\label{e:}
\begin{align}
u_j(\Og) &= \frac{\kk{j}/2-i\Og}{\Delta_j^2 + \left(\kk{j}/2-i\Og\right)^2},\\
v_j(\Og) &=\frac{-\Delta_j}{\Delta_j^2 + \left(\kk{j}/2-i\Og\right)^2}
\end{align}
\end{subequations}
and the effective mechanical susceptibility modified by the dynamical backaction of the two optical modes
\begin{eqnarray}
\susc{eff}^{-1}(\Og) &= \susc{m}^{-1}(\Og) - \sum_j 4 g_j^2 v_j(\Og),
\end{eqnarray}
\end{linenomath*}
where $\susc{m}(\Og) = \Om\left(\Om^2-\Og^2-i\Gm\Og\right)^{-1}$ is the bare mechanical susceptibility.
\\
The quadratures of the field leaking through the most lossy port of our setup (the right ones) are
\begin{linenomath*}
\begin{subequations}
\begin{align}
\hat{X}_j^{\mathrm{out}, R}(\Og) = &-\sqrt{ \kk{j, R}\kk{j, L}}\left(u_j(\Og)\hat{X}^\mathrm{in,L}_{j,\phi_j}  + v_j(\Og) \hat{Y}^\mathrm{in,L}_{j,\phi_j}  \right) \nonumber\\
& - \kk{j, R}\left((u_j(\Og)-1/\kk{j, R})\hat{X}^\mathrm{in, R}_{j}  + v_j(\Og) \hat{Y}^\mathrm{in,R}_{j}  \right) - 2 g_j \sqrt{\kk{j, R}} v_j(\Og) \q(\Og),\\
\hat{Y}_j^{\mathrm{out},R}(\Og) =  &-\sqrt{ \kk{j,R} \kk{j, L} }\left(-v_j(\Og)\hat{X}^\mathrm{in,L}_{j,\phi_j}  + u_j(\Og) \hat{Y}^\mathrm{in,L}_{j,\phi_j}  \right) \nonumber\\
& - \kk{j, R}\left(-v_j(\Og)\hat{X}^\mathrm{in,R}_j  + (u_j(\Og)-1/\kk{j,R}) \hat{Y}^\mathrm{in,R}_j  \right) - 2 g_j \sqrt{\kk{j,R}} u_j(\Og) \q(\Og)
\end{align}
\end{subequations}
\end{linenomath*}

For each mode, we employ a balanced homodyne detector to measure these quadratures.
By varying the phase $\theta_j$ of the local oscillator, we can detect quadratures in a rotated basis:
\begin{linenomath*}
\begin{eqnarray}\label{eqn:homodyne_photocurrent_fourier}
\hat{X}_j^{\theta_j}(\Og) \equiv \hat{X}_j^{\mathrm{out}, R}(\Og) \cos(\theta_j)+\hat{Y}_j^{\mathrm{out}, R}(\Og) \sin(\theta_j).
\end{eqnarray}
\end{linenomath*}
In order to verify the presence of entangled quadratures pairs between the two modes at some frequency $\Og$, we employ a common DGCZ criterion. 
In the frequency domain, this requires calculating symmetrized power spectral densities (PSD) and cross PSD between the modes. 
These PSDs are generally expressed by
\begin{linenomath*}
\begin{equation}\label{e:PSD_general}
\overline{S}^\mathrm{out}_{\hat{X}_j^{\theta_j} \hat{X}_k^{\theta_k}}(\Og) = \frac{1}{2}\delta_{jk} + f^\mathrm{imp}_{jk}(\Og) \Sqq(\Og) + \overline{S}^\mathrm{cor}_{jk}(\Og),
\end{equation}
\end{linenomath*}
where $ \frac{1}{2}\delta_{jk}$ is the shot noise homodyne imprecision,
\begin{linenomath*}
\begin{eqnarray}\label{e:psd_q}
\Sqq(\Og) = \left|\susc{eff}(\Og)\right|^2\left(2\Gamma_A^\mathrm{qba}+2\Gamma_B^\mathrm{qba} + 2\G{m}(\bnth + 1/2)  \right)
\end{eqnarray}
\end{linenomath*}
is the PSD of the mechanical displacement driven by the Brownian force and by the quantum backaction forces $\Gamma_j^\mathrm{qba} = g_j^2\kk{j}\left(|u_j(\Og)|^2+|v_j(\Og)|^2\right)$ of the two lasers,
\begin{linenomath*}
\begin{eqnarray}\label{e:f_imp}
f^\mathrm{imp}_{jk}(\Og) = \frac{\sqrt{\Gamma^\mathrm{meas}_j \Gamma^\mathrm{meas}_k}}{4}\mathrm{Re}\left[e^{-i(\theta_j-\theta_k)}c_{jk}(\Og)-e^{-i(\theta_j+\theta_k)}\alpha_{jk}(\Og)\right]
\end{eqnarray}
\end{linenomath*}
is the transduction function between displacement and detected quadrature and
\begin{linenomath*}
\begin{eqnarray}\label{e:s_corr}
\overline{S}^\mathrm{cor}_{jk}(\Og) = -\frac{\sqrt{\Gamma^\mathrm{meas}_j \Gamma^\mathrm{meas}_k}}{4}\left(\mathrm{Re}\left[\susc{eff}(\Og)\right]\mathrm{Im}\left[e^{-i(\theta_j+\theta_k)}\alpha_{jk}(\Og)\right]\right.\\\nonumber
\left.+ \mathrm{Im}\left[\susc{eff}(\Og)\right]\mathrm{Re}\left[e^{-i(\theta_j-\theta_k)}\beta_{jk}(\Og)\right] \right)
\end{eqnarray}
\end{linenomath*}
is the correlation between the shot noise and the displacement fluctuations induced by the quantum backaction noise.
We also introduced, for the sake of simplicity, the following definitions
\begin{linenomath*}
\begin{subequations}
\begin{align}
\alpha_{jk}(\Og) &= \kk{j}\kk{k}\left(\chi_{\mathrm{c},j} (\Og)\chi_{\mathrm{c},k}(-\Og)+\chi_{\mathrm{c},k}(\Og)\chi_{\mathrm{c},j}(-\Og)\right)=\alpha_{kj}(\Og),\\
\beta_{jk}(\Og) &= \kk{j}\kk{k}\left(\chi_{\mathrm{c},j}(\Og)\chi_{\mathrm{c},k}(\Og)^*-\chi_{\mathrm{c},j}(-\Og)\chi_{\mathrm{c},k}(-\Og)^*\right)= \beta_{kj}(\Og)^*,\\
c_{jk}(\Og) &= \kk{j}\kk{k}\left(\chi_{\mathrm{c},j}(\Og)\chi_{\mathrm{c},k}(\Og)^*+\chi_{\mathrm{c},j}(-\Og)\chi_{\mathrm{c},k}(-\Og)^*\right)= c_{kj}(\Og)^*,\\
\Gamma^\mathrm{meas}_{j} &= \frac{4g_j^2}{\kk{j}}\,\eta_{j}\eta_{j}^\mathrm{c},
\end{align}
\end{subequations}
\end{linenomath*}
where $\chi_{\mathrm{c},j}(\Og) = u_j(\Og)-i v_j(\Og)$ is the cavity field susceptibility, $\eta_{j}^\mathrm{c} = \kk{j, R}/\kk{j}$ the cavity overcoupling, $\eta_j$ the detection efficiency and $\Gamma^\mathrm{meas}_{j}$ is the measurement rate.
The DGCZ criterion for entanglement is based on a pair of EPR observables like $\hat{X}_\pm(\Og)=\hat{X}_A^{\theta_A}(\Og) \pm \hat{X}_B^{\theta_B}(\Og)$ and $\hat{Y}_\pm(\Og)=\hat{Y}_A^{\theta_A}(\Og) \pm \hat{Y}_B^{\theta_B}(\Og)$.
The state is entangled if the inseparability $\mathcal{I}(\Og)<1$, where
\begin{linenomath*}
\begin{eqnarray}\label{e:DGCZ_spectrum}
\mathcal{I}(\Og)\equiv\frac{\overline{S}_{\hat{X}_+\hat{X}_+}(\Og) + \overline{S}_{\hat{Y}_-\hat{Y}_-}(\Og)}{2}=1 + f_{q\mathcal{I}}(\Og)\Sqq(\Og) + \mathcal{I}_\mathrm{cor}(\Og),
\end{eqnarray}
\end{linenomath*}
with
\begin{linenomath*}
\begin{subequations}
\begin{align}
f_{q\mathcal{I}}(\Og) &=\frac{1}{4}\mathrm{Re}\left[ \Gamma^\mathrm{meas}_Ac_{AA}(\Og)+\Gamma^\mathrm{meas}_Bc_{BB}(\Og) + 2\sqrt{\Gamma^\mathrm{meas}_A\Gamma^\mathrm{meas}_B}\alpha_{AB}(\Og)e^{-\imath 2\Theta}\right],\\
\mathcal{I}_\mathrm{cor}(\Og) &= -\mathrm{Im}[\susc{eff}(\Og)]\mathrm{Re}\left[\frac{\Gamma^\mathrm{meas}_A\beta_{AA}(\Og)+\Gamma^\mathrm{meas}_B\beta_{BB}(\Og)}{4}\right] + \mathrm{Re}[\susc{eff}(\Og)]\mathrm{Im}\left[\frac{\sqrt{\Gamma^\mathrm{meas}_A\Gamma^\mathrm{meas}_B}\alpha_{AB}(\Og)e^{-\imath2\Theta}}{2}\right].
\end{align}
\end{subequations}
\end{linenomath*}
and $\Theta = (\theta_A+\theta_B)/2$.
\\

\subsection{Toy Model}\label{s:ToyModel}
It is useful and instructive to consider the theory presented above in the limits of unresolved-sideband cavity ($\kk{j}\gg\Om, \Delta_j$), resonant drive lasers ($\Delta_j=0$) and identical measurement rates ($\Gamma^\mathrm{meas}_{A}=\Gamma^\mathrm{meas}_{B}=\Gamma_\mathrm{meas}$). These limits, apart from describing clearly the underlying physics, are also well suited for the experiments described in the main text.
\\
In fact, within these limits, the joint EPR spectrum simplifies to
\begin{linenomath*}
\begin{multline}\label{e:DGCZ_spectrum_toy}
\mathcal{I}(\Og)\approx 1 + 4 \Gamma_\mathrm{meas}|\chim(\Og)|^2\left(2\Gamma_A^\mathrm{qba}+2\Gamma_B^\mathrm{qba} + 2\G{m}(\bnth+1/2)\right)\left(1+\cos(2\Theta)\right) \\ - 4\Gamma_\mathrm{meas}\mathrm{Re}[\chim(\Og)]\sin(2\Theta),
\end{multline}
\end{linenomath*}
where the quantum backaction rate, in these limits, takes the simple form $\Gamma_j^\mathrm{qba}=4g_j^2/\kk{j}$. 
Whenever, at any frequency $\Og$, the value of such spectrum is below the quantum vacuum level $1$, the quadrature fluctuations at $\Og$ are entangled. 
We have already noticed, in the main text, the close analogy between the physics described by Equation~\eqref{s:ToyModel} and the case of single continuous field ponderomotive squeezing. 
A more detailed look at this connection is presented in Sec. \ref{s:JointSqueezing}.
Note that the correlations term, which is responsible for the entanglement, always vanishes at the mechanical resonance frequency. 
One can minimize Equation~\eqref{e:DGCZ_spectrum_toy} over the frequency $\Og$ and the detection angle $\Theta$ to show that its lower bound is
\begin{eqnarray}
\mathcal{I}(\Og)\ge1-\frac{\eta_\mathrm{meas}}{2},
\end{eqnarray}
where $\eta_\mathrm{meas}$ is the {\it total} measurement efficiency defined as 
\begin{equation}
\eta_\mathrm{meas} \equiv\frac{\sum_j \Gamma^\mathrm{meas}_j}{\sum_j \Gamma_j^\mathrm{qba} + \Gm\left(\bnth+1/2\right)} = \frac{2\Gamma_\mathrm{meas}}{\Gamma_A^\mathrm{qba}+\Gamma_B^\mathrm{qba}+ \Gm\left(\bnth+1/2\right)}.
\end{equation}
\\
\subsection{Covariance matrix}
All the states described above are Gaussian states, due to the quadratic nature of the optomechanical interaction and the assumption of Gaussian white noise at the input. 
Such a state can be fully described by its covariance matrix $\boldsymbol{\sigma}$, i.e. a symmetric matrix containing the correlations between systems' quadratures:
$\sigma_{ij} = \langle x_i x_j + x_j x_i \rangle/2-\langle x_i \rangle \langle x_j\rangle$, where the state vector 
$\mathbf{x} = \left(\hat{X}_A,\hat{Y}_A,\hat{X}_B,\hat{Y}_B\right)^\mathrm{T}$. 
Hereby, we consider the covariance matrix of the state right before the detector, i.e. {\it including} all the optical losses.
\\
This covariance matrix can be described, in block form, by three $2\times2$ matrices: 
\begin{eqnarray}\label{e:CovMat}
    \boldsymbol{\sigma}=
    \begin{pmatrix}
    \mathbf{\alpha} & \mathbf{\gamma}\\
    \mathbf{\gamma}^T & \mathbf{\beta}
    \end{pmatrix}.
\end{eqnarray}
Two of these submatrices,  $\mathbf{\alpha}$ and $\mathbf{\beta}$, describe the individual subsystems while the third, $\mathbf{\gamma}$, describes the correlations between these subsystems,

A sufficient criterion for such state to be entangled is that $2\tilde{\nu}_-<1$, where $\tilde{\nu}_-$ is the lowest symplectic eigenvalue of the partial transposed covariance matrix\cite{adesso2004extremal}. 
Given the matrix in Equation~\eqref{e:CovMat}, the minimum symplectic eigenvalues $\tilde{\nu}_-$ is:
\begin{eqnarray}
    \tilde{\nu}_-=\sqrt{\frac{\Delta(\boldsymbol{\sigma})-\sqrt{\Delta(\boldsymbol{\sigma})^2-4\mathrm{Det}\boldsymbol{\sigma}}}{2}},
\end{eqnarray}
where $\Delta(\boldsymbol{\sigma})=\mathrm{Det}\mathbf{\alpha}+\mathrm{Det}\mathbf{\beta}-2\mathrm{Det}\mathbf{\gamma}$. 
The physical meaning of $\tilde{\nu}_-$ can be understood as\cite{zippilli2015entanglement} $2\tilde{\nu}_-=\mathrm{min}\,\mathcal{I}(\Og)$, where the inseparability is minimized over all local linear unitary Bogolioubov operations, such as rotations and squeezing. 
Thus, $\tilde{\nu}_-$ sets a lower bound of the DGCZ inseparability at arbitrary quadratures. 
The minimum symplectic eigenvalue of the partial transposed covariance matrix is related to well-known entanglement measure logarithmic negativity $E_N$ as $E_N=\mathrm{max}\left[0,-\mathrm{ln}\,2\tilde{\nu}_-\right]$.
\\
For the toy model defined above, the covariance matrix can be written as:
\begin{eqnarray}
\mathbf{V}=
\begin{pmatrix}
\frac{1}{2}                 &        2\Gamma_\mathrm{meas}\mathrm{Re}[\chi_m(\Og)]                         &        0                  &   2\Gamma_\mathrm{meas}\mathrm{Re}[\chi_m(\Og)] \\
2\Gamma_\mathrm{meas}\mathrm{Re}[\chi_m(\Og)]  &   \frac{1}{2}+8\Gamma_\mathrm{meas}|\chi_m(\Og)|^2\Gdec  &   2\Gamma_\mathrm{meas}\mathrm{Re}[\chi_m(\Og)]  &   8\Gamma_\mathrm{meas}|\chi_m(\Og)|^2\Gdec \\
0                       &        2\Gamma_\mathrm{meas}\mathrm{Re}[\chi_m(\Og)]                         &        \frac{1}{2}                  &   2\Gamma_\mathrm{meas}\mathrm{Re}[\chi_m(\Og)] \\
2\Gamma_\mathrm{meas}\mathrm{Re}[\chi_m(\Og)]  &   8\Gamma_\mathrm{meas}|\chi_m(\Og)|^2\Gdec  &   2\Gamma_\mathrm{meas}\mathrm{Re}[\chi_m(\Og)]  &  \frac{1}{2}+8\Gamma_\mathrm{meas}|\chi_m(\Og)|^2\Gdec
\end{pmatrix},
\end{eqnarray}
where 
$\Gdec=\Gamma_A^\mathrm{qba}+\Gamma_B^\mathrm{qba}+\Gm(\bnth+1/2)$. The minimum symplectic eigenvalue can be promptly calculated to be
\begin{equation}
    2\tilde{\nu}_-(\Og) = \sqrt{1+16\Gamma_\mathrm{meas}|\chim(\Og)|^2\Gdec\left(1-\sqrt{1+\frac{\mathrm{Re}[\chim(\Og)]^2}{4|\chim(\Og)|^4\Gdec^2}}\right)}.
\end{equation}
If we focus on the vicinity of mechanical frequency ($|\delta_\mathrm{m}|=|\Om-\Og|\ll\Om$), the ratio in the inner square root $\mathrm{Re}[\chim(\Og)]^2/4|\chim(\Og)|^4\Gdec^2\approx\delta_\mathrm{m}^2/\Gdec^2$. In the limit of $\Gdec\gg\delta_\mathrm{m}$, we can take the approximation $\sqrt{1+\delta_\mathrm{m}^2/\Gdec^2}\approx1+\delta_\mathrm{m}^2/(2\Gdec^2)$. At the same time, in the limit of $\delta_\mathrm{m}\gg\Gm$, $4|\chim(\Og)|^2\approx1/\delta_\mathrm{m}^2$. Then the expression can be simplified to:
\begin{equation}
    2\tilde{\nu}_-(\Og) \approx \sqrt{1-\frac{2\Gamma_\mathrm{meas}}{\Gdec}}=\sqrt{1-\etam}.
\end{equation}
Thus, in the limit of very strong measurement strength $\etam\rightarrow1$ one can achieve arbitrarily strong entanglement for optical modes at frequency $\delta_\mathrm{m}$, where $\Gm\ll\delta_\mathrm{m}\ll\Gdec$. 
Though the last expression is frequency independent, deviation from the limits results in the frequency dependence feature of Figure~4 in main text.
%
%
%
%
%
%
%
%
\subsection{Entanglement as Joint Quadrature Squeezing}\label{s:JointSqueezing}
Here, we comment further on the toy model presented in Sec. \ref{s:ToyModel}, to highlight the form of correlations generated in this system.  
First, recall that, in the toy model, the dynamics of the optical modes are given by:
\begin{linenomath*}
\begin{subequations}
  \begin{align}
    \label{eq:toy_model_eom}
    \hat{X}_j^\mathrm{out} &= -\hat{X}_j^\mathrm{in},\\
    \hat{Y}_j^\mathrm{out} &= -\hat{Y}_j^\mathrm{in}-2\sqrt{\Gamma_\mathrm{qba}}\chi_\mathrm{m}(t)*\left( \sqrt{2\Gamma_\mathrm{m}}\hat{P}_\mathrm{in} + \sqrt{4\Gamma_\mathrm{qba}}\hat{X}_+^\mathrm{in} \right),
  \end{align}
\end{subequations}
\end{linenomath*}
where we also assume, for the sake of clarity, equal cavity linewidths $\kk{}=\kk{A}=\kk{B}$
\begin{figure*}[!bb]
\begin{center}
\includegraphics[width=0.8\columnwidth]{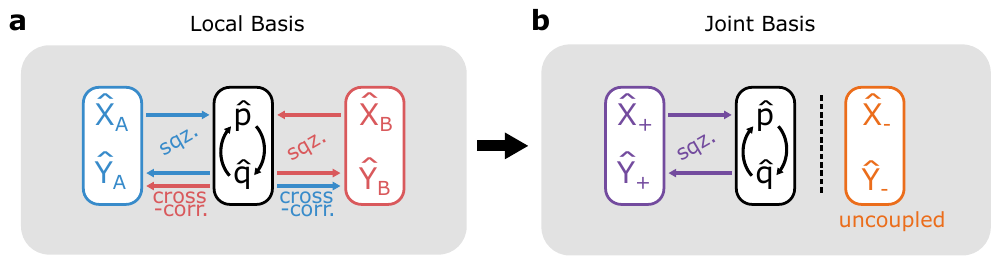}
\caption{{\bf Dynamics in the Joint Mode Basis.}
{\bf a}, Schematic illustration of the couplings between the optical quadratures ($\hat{X}_A,\hat{Y}_A,\hat{X}_B,\hat{Y}_B$) and mechanical position/momentum ($q,p$).  Each laser generates self-squeezing as well as cross-correlations. {\bf b}, Schematic illustration of the couplings after moving to the (non-local) joint basis ($\hat{X}_{\pm}=\hat{X}_A\pm\hat{X}_B,\hat{Y}_{\pm}=\hat{Y}_A\pm\hat{Y}_B$).  Here, there is only a self-squeezing of the sum-mode, while the difference-mode remains uncoupled from all system dynamics. 
}
\label{f:si-couplings}
\end{center}
\end{figure*}

The links between the different optical and mechanical quadratures are illustrated in Figure \ref{f:si-couplings}(a).  
As mentioned in the main text, there is self-squeezing as well as cross-correlations, both generated by amplitude fluctuations driving motion, which then drives phase fluctuations.  
If we move to the basis of the joint sum and difference quadratures, $\hat{X}_{\pm}=\hat{X}_A\pm\hat{X}_B$ and $\hat{Y}_{\pm}=\hat{Y}_A\pm\hat{Y}_B$ (applied to both input and output fields), we can rewrite the dynamics as follows:
\begin{linenomath*}
\begin{subequations}
  \begin{align}
    \label{eq:EPR_variables}
    \hat{X}_+^\mathrm{out} &= -\hat{X}_+^\mathrm{in},\\
    \hat{Y}_+^\mathrm{out}&= -\hat{Y}_+^\mathrm{in}-4\sqrt{\Gamma_\mathrm{qba}}\chi_\mathrm{m}(t)*\left( \sqrt{2\Gamma_\mathrm{m}}\hat{P}_\mathrm{in} + \sqrt{4\Gamma_\mathrm{qba}}\hat{X}_+^\mathrm{in} \right),\\
    \hat{X}_-^\mathrm{out} &= -\hat{X}_-^\mathrm{in},\\
    \hat{Y}_-^\mathrm{out} &= -\hat{Y}_-^\mathrm{in}.
  \end{align}
\end{subequations}
\end{linenomath*}
Note that the difference mode becomes completely de-coupled from the optomechanical dynamics, simply remaining in its initial vacuum state.  
Meanwhile, the sum mode undergoes the usual optomechanical interaction, generating self-squeezing through the mechanical motion.  
This squeezing is twice as strong as the self-squeezing of either individual laser. 
\begin{figure*}[!th]
\begin{center}
\includegraphics[width=1.0\columnwidth]{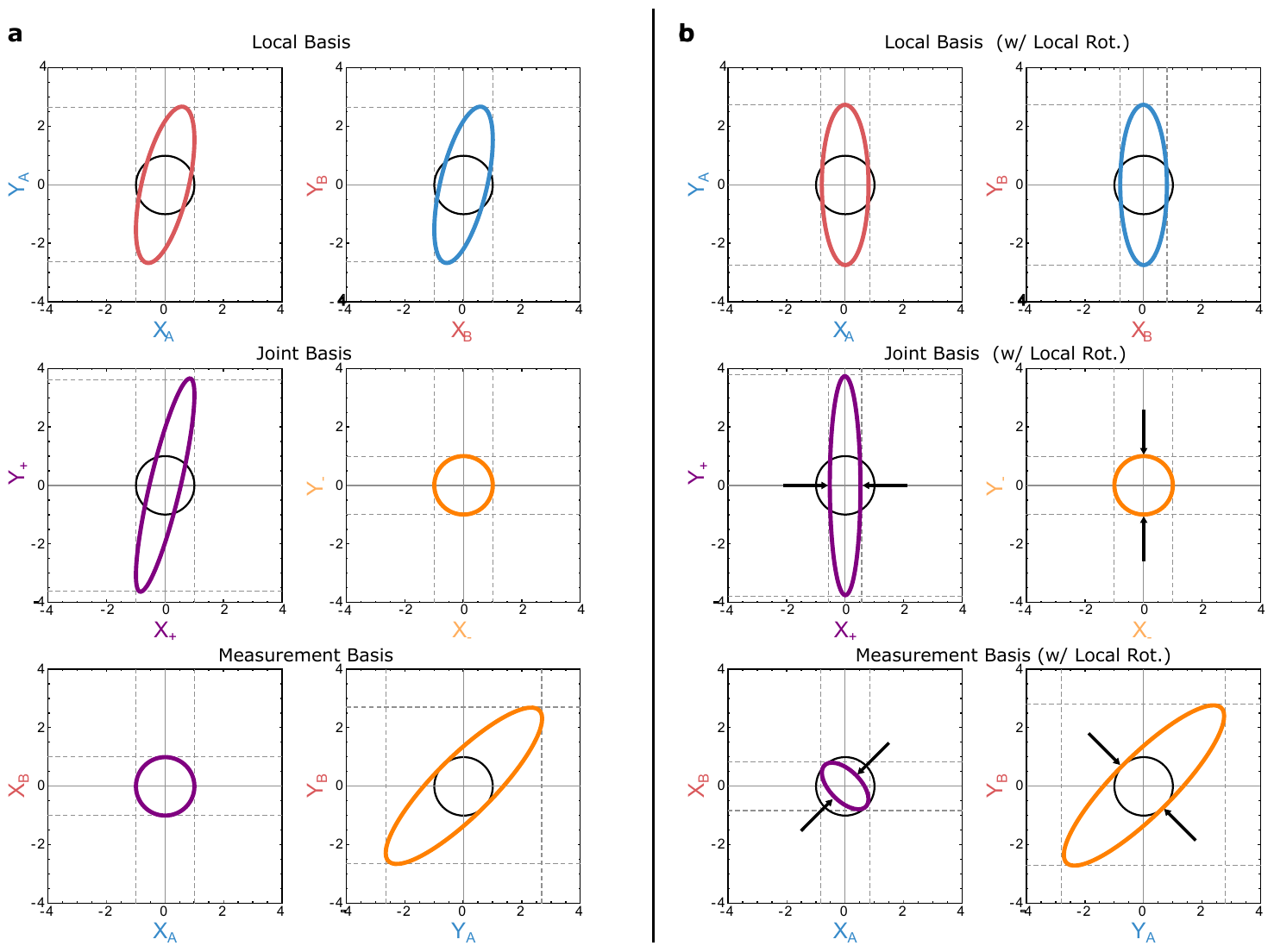}
\caption{{\bf Entanglement in Phase Space.}
{\bf a}, Different phase space portraits of the system correlations (i.e. different cuts of the 4-dimensional covariance ellipsoid). In all subplots, the black circle indicates vacuum noise, and the dashed lines indicate the marginal variances along each axis.  The top row is in the (local) cavity quadrature basis, where each system simply exhibits self-squeezing. The middle row moves to the joint bases, where we see that the difference mode remain in vacuum, while the sum-mode displays self-squeezing (with a non-zero squeezing angle).  We note that the squeezing of this joint quadrature is stronger than the sub-system self-squeezing. The bottom row corresponds to the ellipses we are able to directly measure in our system (i.e. simultaneous $\{\hat{X}_A,\hat{X}_B\}$ or $\{\hat{Y}_A,\hat{Y}_B\}$). {\bf b},  Similar noise ellipses as in {\bf a}, but with a local rotation of the $A$,$B$ subsystems (i.e. $\theta_A=\theta_B\neq0$). As a result, the joint-basis ellipses (middle row) are aligned such that we now see $V(X_+)+V(Y_-)$ will violate the DGCZ criterion, as indicated by the black arrows.  In our measurements, these variances are extracted from the diagonal/anti-diagonal variances (bottom row).  We note that the bottom row closely matches the experimental noise ellipses of Figure~2 in the main text.  
}
\label{f:si-ellipses}
\end{center}
\end{figure*}

Figure~\ref{f:si-ellipses} shows covariance ellipses based on this model, illustrating how the correlations appear in either basis. (Note that in the figure and discussion below, we refer only to output modes, but drop the $^\mathrm{out}$ superscript for convenience.) 
We note that the local-basis ellipses are insufficient to characterize the system, since there are unseen correlations ($Y_A$ and $Y_B$, for example).  
On the other hand, since the joint quadrature bases are decoupled, these covariance ellipses completely describe all significant system correlations.  

To link these covariance ellipses to entanglement, recall that the DGCZ criterion requires that \\ $\left(V(X_+)+V(Y_-)\right)/2<1$. 
For the above quadrature definitions, this is not yet true, since the squeezing of the sum mode occurs for some non-zero squeezing angle (see Figure~\ref{f:si-ellipses}(a)).  
However, by applying a local rotation to each laser (i.e. selecting $\theta_A=\theta_B\neq0$), we can align this sum-mode-squeezing with the basis, such that we find $V(X_+)<1$, while $V(Y_-)$ remains at the vacuum level.  
Thus, we see how self-squeezing of this non-local mode establishes the entanglement of the two fields.
%
%
%
%
%
%
%
%
%
\subsection{Output spectral modes and homodyne detection}
So far, we have been discussing about bipartite entanglement, i.e. entanglement
between two modes. In general, the dynamics described before involves four
\textit{spectral} modes. We show now that, with some assumptions on the
underlying quantum state and homodyne measurements, this quadripartite state can
effectively be reduced to a simpler bipartite state\cite{barbosa2013pra}.

A single, continuous field
$\hat{a}(t)=\int\frac{d\Og}{2\pi}e^{-\imath\Og t}\hat{a}_\Og$
can be seen as a highly multimode system, containing an ensemble of spectral
modes $\hat{a}_\Og$. The bandwidth of this ensemble is defined by the details of
the system which generates the field.
The optomechanical interaction in
Equation~\eqref{e:qle_time} simultaneously affects the
\textit{sideband} modes (a pair of spectral modes symmetric
around the carrier frequency) $\hat{a}_{\Og_L\pm\Om}$. To keep the notation
simple and concise, we drop, in the subscripts, the carrier frequency $\Og_L$ in the following.

\begin{figure}[ht!]
  \centering
  \includegraphics[scale=1]{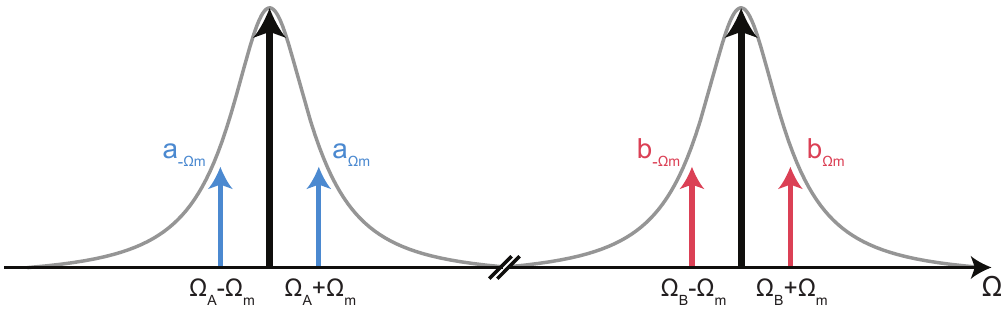}
  \caption{{\bf Output spectral modes involved.} Each laser field drives,
    resonantly, an optical cavity mode and interact with a common mechanical
    mode of frequency $\Om$. As a result, the output propagating fields posses
    non-trivial sideband modes at the mechanical frequency $\Om$. Thus, their
    dynamics is determined by a quadripartite state.}
  \label{fig:output_modes}
\end{figure}

In our case, a multimode optomechanical system is driven by two different
fields, $\hat{a}(t)$ and
$\hat{b}(t)$, which interact with the same mechanical mode. Thus, the interaction
involves four spectral modes, i.e. two pairs of sideband modes
$\hat{a}_{\pm\Om}$ and $\hat{b}_{\pm\Om}$, each around its own carrier
frequency, as shown in Figure~\ref{fig:output_modes}.

Subsequently, each field is measured via spectral homodyne detection, which
corresponds to a Fourier analysis of the homodyne photocurrent, in order to
resolve individual spectral modes. This results in a mixing of the upper and
lower sideband at the analysis frequency $\Og$. For a single field, given the photocurrent
$\hat{i}(t) = e^{-\imath\theta}\hat{a}(t)+ e^{\imath\theta}\hat{a}(t)^\dagger$,
its Fourier component at frequency $\Og$ is
\begin{equation}
  \label{eq:homo_fourier}
  \hat{i}_\Og = \frac{e^{-\imath\theta}\hat{a}_\Og + e^{\imath\theta}(\hat{a}_{-\Og})^\dagger}{\sqrt{2}} = \cos\left( \theta \right)\frac{ \hat{X}_\mathrm{s}+\imath \hat{Y}_\mathrm{a}}{\sqrt{2}} + \sin\left( \theta \right)\frac{ \hat{Y}_\mathrm{s}-\imath \hat{X}_\mathrm{a}}{\sqrt{2}},
\end{equation}
where $\hat{X}_\mathrm{s(a)},\,\hat{Y}_\mathrm{s(a)}$ are the quadratures of the
symmetric (antisymmetric) modes, defined as
$\hat{a}_\mathrm{s(a)}=(\hat{a}_\Og\pm\hat{a}_{-\Og})/\sqrt{2}$.
Equation~\eqref{eq:homo_fourier} shows us that spectral homodyne detection is
truly a two-mode measurement, where the two modes involved are the upper and
lower sidebands $\hat{a}_{\pm\Og}$ or, equivalently, the symmetric and
antisymmetric modes.
When considering the case with two fields, the combined homodyne measurements
directly sample the quadripartite state formed by the spectral modes
$(\hat{a}_{\Og}, \hat{a}_{-\Og}, \hat{b}_\Og, \hat{b}_{-\Og})$.

Generally speaking, such state cannot be completed reconstructed from homodyne
detection, as this measurement is ``blind'' to several correlation terms\cite{barbosa2013pra}.
However, an important special case (valid also for our system) is played by
stationary states with no asymmetry between the energy of sideband modes.
Within these
assumptions, the quadripartite state simplifies to a bipartite one
formed only by the symmetric mode of the two fields.
We also notice that, in this case, the
antisymmetric mode possesses the same quantum statistics as the symmetric one,
apart from a local (then irrelevant for entanglement) rotation in phase space.

This would suggest than a single-mode measurement description for spectral
homodyne detection is possible.
However, the measured quadratures, e.g.
Equation~\eqref{eqn:homodyne_photocurrent_fourier}, are not proper single-mode
quadrature operators (in fact, orthogonal quadratures commute, i.e.
$[\hat{X}_\Og, \hat{Y}_\Og]=0$) but rather semiclassical ones.
Nevertheless, for stationary quantum states, they behave as effective
single-mode quadrature operators as far as second-order moments are concerned,
when the right prescription to calculate noise power is used.
Then, from their statistics, one can reconstruct the covariance matrix for the
symmetric modes.
It is the entanglement of these symmetric modes, belonging to the two fields,
that we observe and report in the main text\cite{barbosa2013pra}.

We also stress that this represents a different situation compared to red-blue
sideband entanglement generated by a resolved-sideband optomechanical
system\cite{barzanjeh2019}. In our case, the correlations of interest are
present in the electronic signal at frequency $\Om/(2\pi)\sim\mathcal{O}(1\,\mathrm{ MHz})$, where technical noise can be
suppressed and quantum limited detection achieved. In the resolved sideband
case, instead, the two entangled modes are single spectral modes and not
sideband modes. The correlations, in the homodyne detection signal, are now
present at zero frequency, where technical noise obstructs quantum-limited measurements. This can be alleviated by employing an
heterodyne detector, which, however, would limit the detection efficiency to 50\%.
%
%
%
%
%
%
%
%
\section{Experimental Calibrations and Data Analysis}

Here we introduce some details of experimental calibrations and data analysis. 
We begin by showing the equivalence between a temporal mode defined in the main text and the FFT of the photocurrents. 
Next, we introduce the fitting of the power spectral densities. 
Finally, we discuss the calibration of balanced homodyne detectors, and the systematic errors caused by non-ideal detection. 

\subsection{Mode Decomposition by Fourier Transform}
\begin{figure*}[h]
\begin{center}
\includegraphics[width=.5\columnwidth]{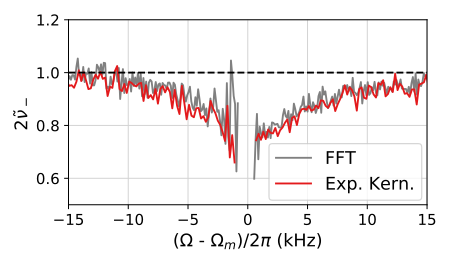}
\caption{{\bf Comparison of temporal mode definitions.}
}
\label{f:FFT}
\end{center}
\end{figure*}

In the main text, we introduce temporal modes (defined in Equation~(3)), obtained by filtering the optical quadratures with a kernel, $h(t)$.  In the data analysis of Figures~2 and 3, we construct such modes using an exponential kernel (in practice, applying a digital, 4th order butterworth filter to the mixed-down signal).  In Figure 4, we wish to analyze the frequency-dependent statistics of such modes.  For computational speed, we accomplish this by calculating the FFT of the photocurrents, which corresponds to using a rectangular/boxcar kernel in Equation~(3).  While the profile of this FFT-defined mode will differ slightly from the modes of Figs. 2 and 3, we note here that the differences are not significant for our analysis. Figure~\ref{f:FFT} compares, for example, the frequency dependence of $\tilde{\nu}_-$, for modes calculated by Fourier transform and by exponential kernel.

\subsection{Fitting of power spectral densities}
\begin{figure*}
\begin{center}
\includegraphics[scale=1.0]{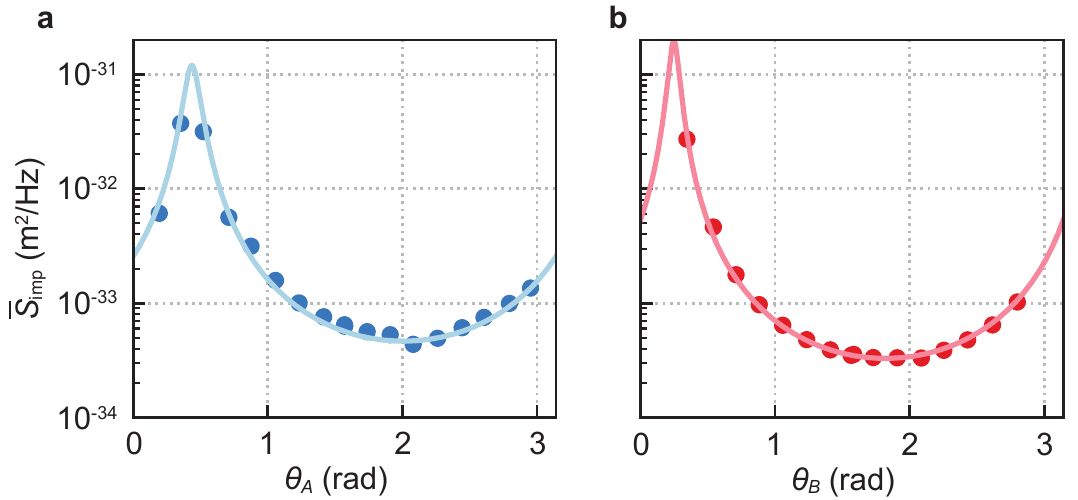}
\caption{{\bf Imprecision for different quadratures.}
{\bf a}, {\bf b}, Displacement imprecision for laser A and B, respectively, as a function of their homodyne angle. The light lines are fits.
}
\label{f:si-imprecision}
\end{center}
\end{figure*}

Here we describe the analysis procedure for fitting the spectra shown in Figure~4 of the main text (the frequency-dependent DGCZ measurements).  
Recall that each measurement run (i.e. each joint basis angle $\Theta$) involves three sequential measurements, each of which consist of two simultaneously acquired photocurrent streams.  These three steps measure (1) shot noise (by blocking cavity outputs), (2) $\{\hat{X}_{A}^{\theta_A},\hat{X}_{B}^{\theta_B}\}$, and (3) $\{\hat{Y}_{A}^{\theta_A},\hat{Y}_{B}^{\theta_B}\}$ (by advancing the homodyne angles from step (2) by $\pi/2$).
We start by fitting the detection efficiency of each system. Following a standard technique\cite{gorodetsky2010aa}, we can use a phase modulation tone in the spectra to calibrate the shot noise background into displacement units. We fit these displacement spectra (at various homodyne angles) to $\overline{S}_j^\mathrm{imp}(\theta_j) = 2 x_\mathrm{ZPF}^2 / f_{jj}^\mathrm{imp}(\overline{\Og})$, where $f_{jj}^\mathrm{imp}(\overline{\Og})$ is defined in Equation~\eqref{e:f_imp} and $\overline{\Og}$ is the fixed Fourier frequency at which we measure the background value (see Figure~\ref{f:si-imprecision}). From these fits we get $\eta_A=60\%$ and $\eta_B=77\%$.  (Note that in this and future fits, the exact values of $\theta_A$ and $\theta_B$ are extracted from DC values of the balanced photocurrents.)

\begin{figure*}
\begin{center}
\includegraphics[scale=1.0]{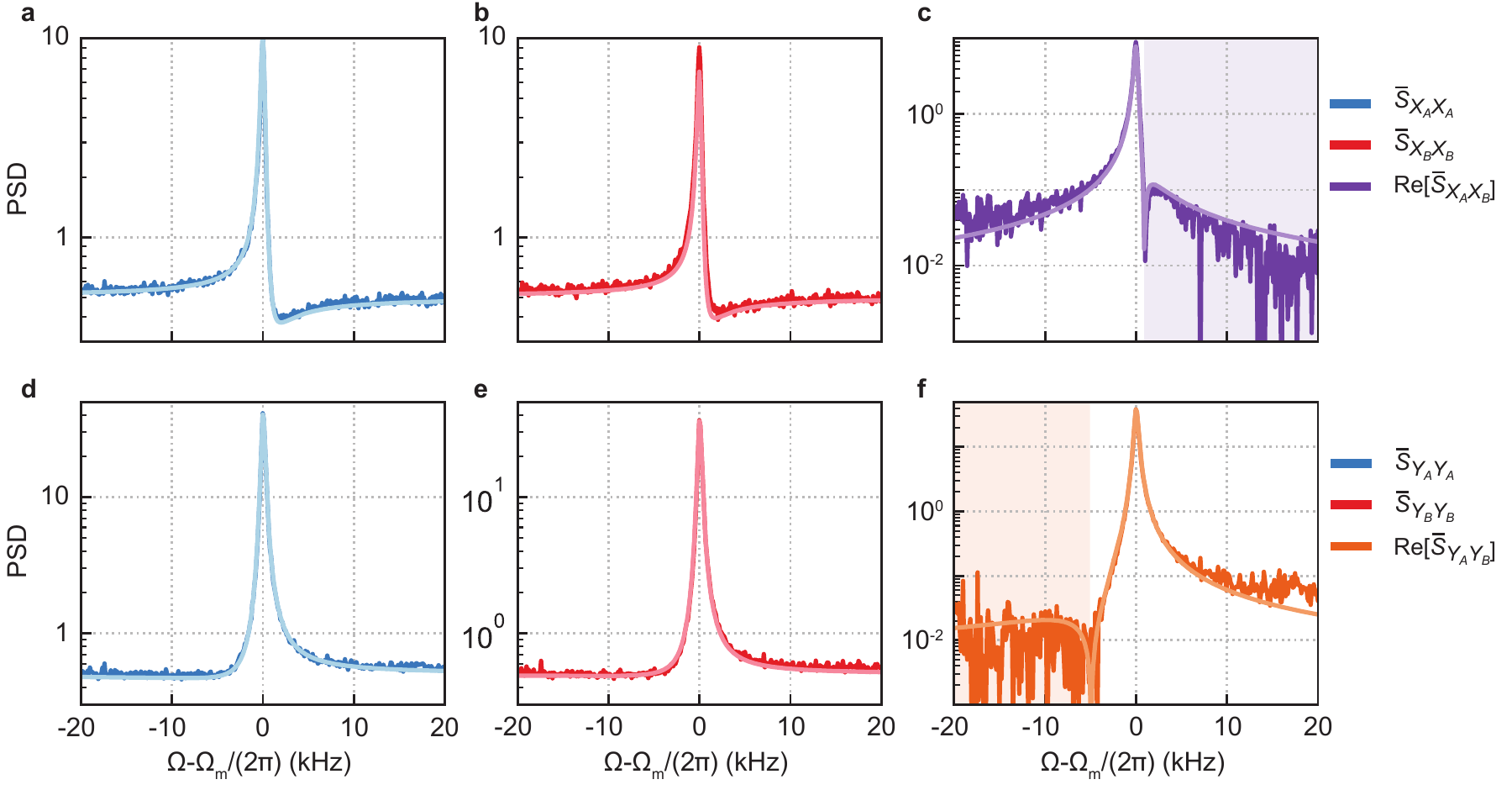}
\caption{{\bf Power spectral density from a measurement run.}
{\bf a}, {\bf b}, {\bf c}, PSDs for the $\hat{X}$ quadratures of the two lasers ($\Theta\approx0$). We show the absolute value of the cross-spectrum in {\bf c} for the sake of visualization and note that the shaded area is actually negative. Light-coloured lines are the result of a simultaneous fit of all 6 spectra.
{\bf d}, {\bf e}, {\bf f}, PSDs for the $\hat{Y}$ quadratures of the two lasers.
}
\label{f:si-spectra}
\end{center}
\end{figure*}

\begin{figure*}
\begin{center}
\includegraphics[scale=1.0]{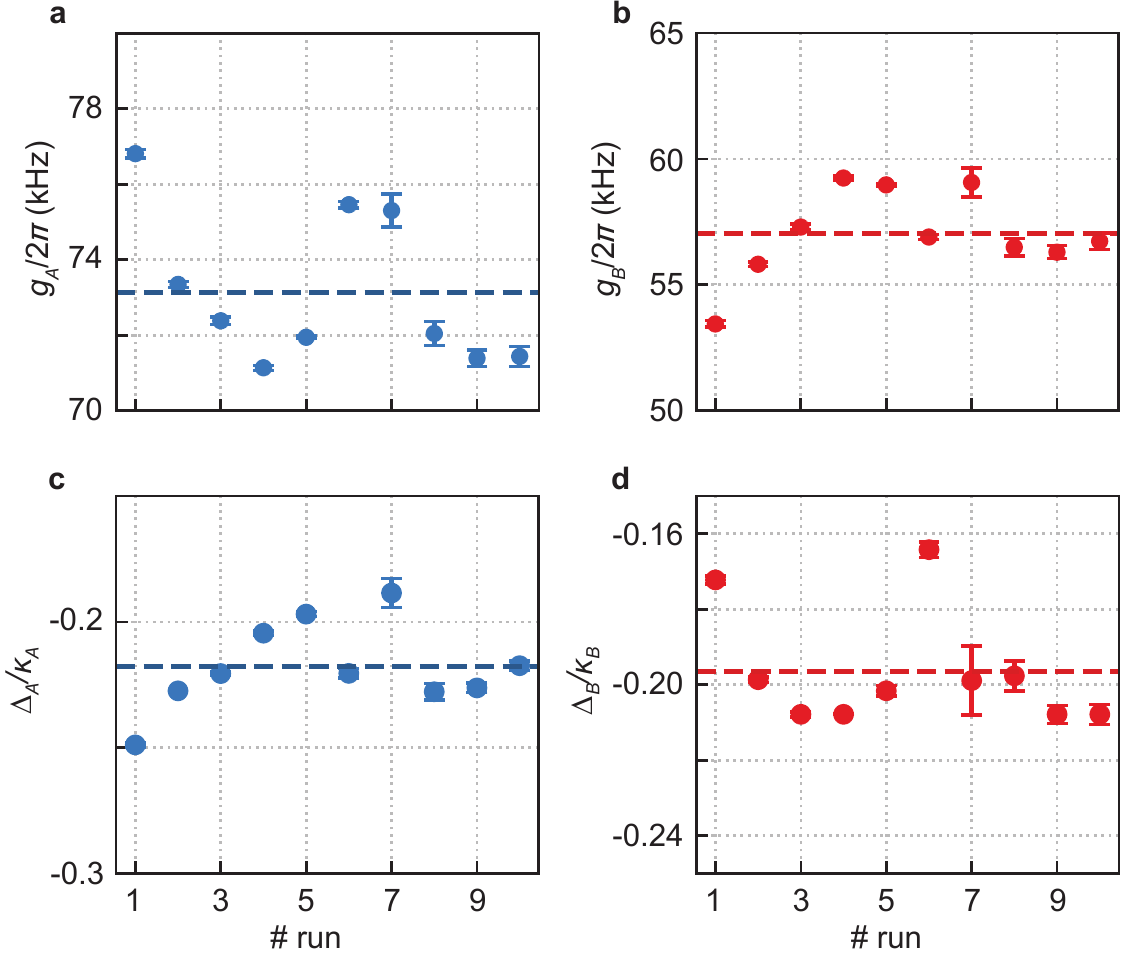}
\caption{{\bf Fitted parameters for all measurement runs.}
{\bf a}, {\bf b}, Fitted optomechanical coupling $g$ for laser A and B, respectively.
{\bf c}, {\bf d}, Fitted detuning $\Delta$ for laser A and B, respectively, in units of the respective cavity linewidth.
The dashed lines are mean values. Error bars represents a single standard deviation coming from the fits.
}
\label{f:si-parameters}
\end{center}
\end{figure*}

We then move to analyze and fit the photocurrent spectra from from steps (2) and (3). 
Each measurement run yields 6 different spectra: $\overline{S}_{\hat{X}_A^{\theta_A} \hat{X}_A^{\theta_A}}$, $\overline{S}_{\hat{X}_B^{\theta_B} \hat{X}_B^{\theta_B}}$, $\mathrm{Re}\left(\overline{S}_{\hat{X}_A^{\theta_A} \hat{X}_B^{\theta_B}}\right)$ and the analogous for $\hat{Y}_j^{\theta_j}$. We normalize them to the shot noise level acquired in step (1). We fit all 6 spectra simultaneously to Equation~\eqref{e:PSD_general} with the appropriate choice of angles and labels $j, k$. The only free parameters in this fit are the coupling $g_j$ and the detuning $\Delta_j$, in order to account for slow drift during the measurements. An example of such fitted spectra is given in Figure~\ref{f:si-spectra}.
Finally, we show in Figure~\ref{f:si-parameters} the fitted parameters for all the measurement runs. Their mean values are used to plot theory lines in Figure~2 and Figure~3 of the main text.
\subsection{Calibration of balanced homodyne detector }\label{s:det_cal}
\begin{figure*}
\begin{center}
\includegraphics[scale=1]{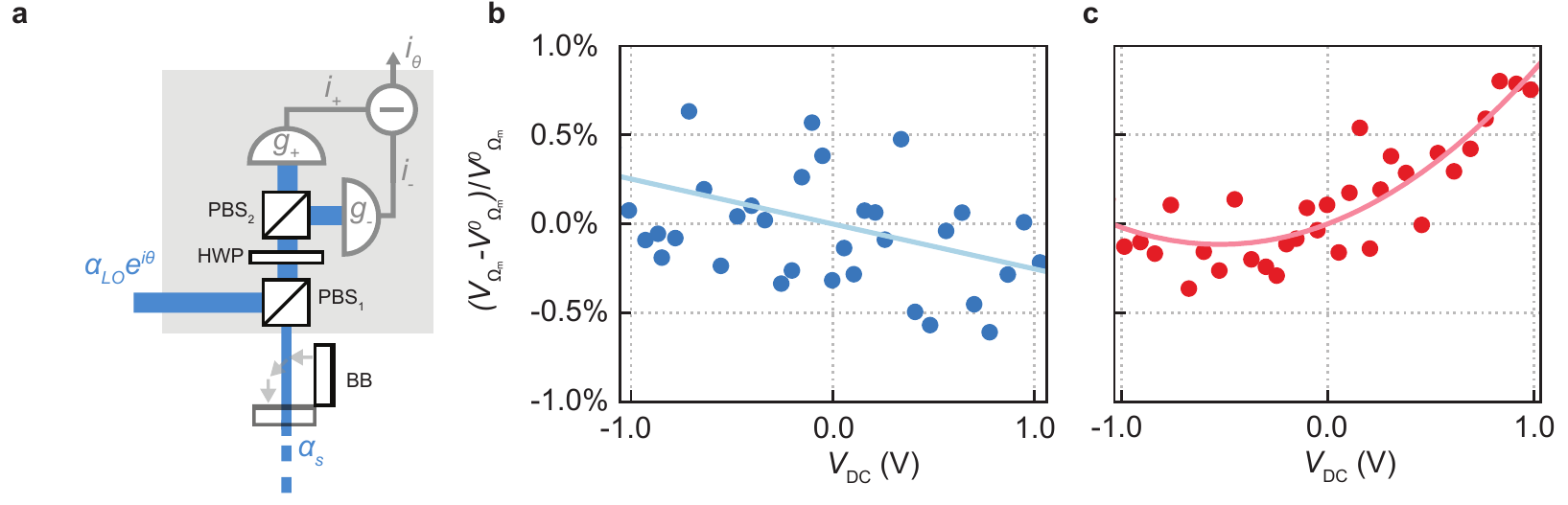}
\caption{{\bf Systematic errors in balanced homodyne detectors.}
{\bf a}, A local oscillator with coherent amplitude $\alpha_\mathrm{LO}e^{\imath\theta}$ is combined with an orthogonally linearly polarized field $\alpha_s$ on a polarizing beam splitter (PBS). A half-wave plate (HWP) rotates the polarization by $45^\circ$ (in a balanced configuration) and a final PBS splits the mixed fields. Two photodiodes (labeled $+$ and $-$) with overall gain $g_\pm$ detect the fields. The resulting photocurrent $i_\pm$ are then subtracted and the differential photocurrent $i$ is given as output. In the experiment, two such balanced homodyne detectors are used (one for each laser).
To calibrate the detector, the signal field can be blocked by rotating the beam block (BB) and the local oscillator power can be unbalanced by rotating the HWP.
{\bf b}, {\bf c}, Difference in measured variance at around $\Om$, relative to the variance when the detector is balanced, as a function of the DC component of the differential photocurrent (here in terms of voltage) for, respectively, laser A and laser B. Every $V_\mathrm{DC}$ corresponds to a rotation of the HWP.  The light blue (red) line is fit to a linear (quadratic) function.
}
\label{f:calib}
\end{center}
\end{figure*}

In general, the main results of the manuscript depend only on comparing measured spectra/variances to a shot noise reference level. 
To measure this reference level, we block the signal, so only the equally-distributed local oscillator reaches the balanced homodyne detector (BHD). 
The conditions of this measurement differ from a generic quadrature measurement, where the BHD is locked at angle $\theta$ and the power on each diode can differ, due to interference between the LO and the coherent signal from the experiment.
In principle, this can lead to small systematic effects related to photodiode gain differences and classical laser noise.
These effects can be characterized by shot noise measurements in which the LO power is unbalanced\cite{safavi2013squeezed}, as we derive below. 
(We note already that no significant systematic artifacts are present in our data, but we present this characterization for completeness.)
We consider optical fields at the photodiodes of a balanced homodyne detector, with real local oscillator field $\alpha_\mathrm{LO}$, real signal field $\alpha_\mathrm{s}$, and classical laser amplitude noise $n(t)$ (see Figure~\ref{f:calib}{\bf a}). The local oscillator power is imbalanced by a (relative) amount $\delta$. The field at diode $\pm$, where $+$ or $-$ refers to the different photodiodes, is given by:
\begin{equation}
    \frac{\alpha_\mathrm{LO}}{\sqrt{2}}\sqrt{1\pm\delta}(1+n(t))e^{-i\theta}\pm\frac{\alpha_\mathrm{s}}{\sqrt{2}}(1+n(t))+\hat{a}_{\pm,\mathrm{vac}},
\end{equation}
where $\theta$ is the relative phase between LO and signal, $\hat{a}_{\pm,\mathrm{vac}}$ is the vacuum noise, and the $\pm$ before $\alpha_\mathrm{s}$ term results from reflection of field from the final beam splitter. 
As the variance of vacuum $\hat{a}_{\pm,\mathrm{vac}}$ is invariant under rotation, the photocurrent spectrum is completely determined by the amplitude of the classical terms.
In a balanced configuration ($\delta=0$), this amplitude is given by
\begin{equation}\label{e:exp}
    (1+n(t))\sqrt{\frac{\alpha_\mathrm{LO}^2}{2}+\frac{\alpha_\mathrm{s}^2}{2}\pm\alpha_\mathrm{LO}\alpha_\mathrm{s}\cos{\theta}} \approx (1+n(t))\sqrt{\frac{\alpha_\mathrm{LO}^2}{2}\pm\alpha_\mathrm{LO}\alpha_\mathrm{s}\cos{\theta}},
\end{equation}
where we use the assumption that  $\alpha_\mathrm{LO}^2\gg\alpha_\mathrm{s}^2$. 
Alternatively, if we block the signal and consider an unbalanced ($\delta\neq0$) detector, the amplitude is given by:
\begin{equation}\label{e:unBHD}
    (1+n(t))\sqrt{\frac{\alpha_\mathrm{LO}^2}{2}(1\pm\delta)}.
\end{equation}
Comparing these two cases, we find that Equations~\eqref{e:exp} and \eqref{e:unBHD} are equivalent when $\delta=2\alpha_\mathrm{s}/\alpha_\mathrm{LO}\cos{\theta}$.  
Thus, all systematic artifacts caused by measuring $\theta\neq\pi/2$, can be completely characterized by blocking the signal ($\alpha_\mathrm{s}=0$) and unbalancing the LO.  
In practice, we don't measure $\delta$ but the amplified DC component of the photocurrent, $V_\mathrm{DC} = \alpha_\mathrm{LO}^2/2\left(g_+-g_- + (g_++g_-)\delta\right)$. 
The power spectral density of the amplified differential photocurrent, $V$ is 
\begin{eqnarray}
    \overline{S}_{VV}(\Og) = \underbrace{g_+g_-\alpha_\mathrm{LO}^2 + \left(g_+-g_-\right)V_\mathrm{DC}}_\text{shot noise} + \underbrace{4\overline{S}_\mathrm{nn}(\Og)V_\mathrm{DC}^2}_\text{classical noise},
\end{eqnarray}
where $g_\pm$ are the photodiode gains and $\overline{S}_\mathrm{nn}(\Og)$ is the power spectral density of the amplitude noise $n(t)$.
We vary $V_\mathrm{DC}$ by changing the splitting ratio $\delta$ and measure the average spectral noise $V_{\Om}$ at around $\Om$. 
In Figure~\ref{f:calib}{\bf b} and {\bf c} we show, respectively, the relative variance $(V_{\Om}-V^0_{\Om})/V^0_{\Om}$ for the BHD of laser A and B, where $V^0_{\Om}$ is the variance at $V_\mathrm{DC}=0$, which approximately corresponds to a balanced detector for small gain difference. 
The variance $V^0_{\Om}$ is also the one used as a reference shot noise for the
main results in the main text. The largest deviation we observe ($<1\%$) is much
less then any DGCZ violations reported here and, thus is insignificant.
Nevertheless, we consider such systematics for the best inseparability value reported in the main text. For this measurement, we operate the detectors both at the minimum of the fringe interference (i.e. $V_\mathrm{DC}\approx-1$~V) and at the center ($V_\mathrm{DC}\approx0$~V). We see that an error of $\sim0.3\%$ arises when measuring the amplitude
quadratures of laser A.\\

\section{System parameters}
In Tab.~\ref{tab:parameters} we collect the main symbols and parameters which have been used in this work.
\begin{table}
\centering
\begin{tabular}{lllll}
    \hline
    {\bf Symbol} & {\bf Definition} & {\bf Name} & {\bf Value} & {\bf Value} \\
      &   &   & {\bf mode $A$} & {\bf mode $B$}  \\\hline
    $\Om$ & & Mechanical resonance frequency &\multicolumn{2}{c}{$2\pi\times1.139~\mathrm{MHz}$} \\
    $\Gamma_\mathrm{m}$ & & Mechanical linewidth &\multicolumn{2}{c}{$2\pi\times1.1~\mathrm{mHz}$} \\
    $Q$ & $\Om/\Gm$ & Mechanical quality factor & \multicolumn{2}{c}{$1.03\times10^9$}\\
  $T$ & & Mechanical bath temperature & \multicolumn{2}{c}{$10\,\mathrm{K}$} \\
  $\bnth$ & $(e^{\hbar\Om/\kappa_B T}-1)^{-1}$ & Thermal phonon occupation & \multicolumn{2}{c}{$1.8\times10^5$} \\
    $\meff$ & & Effective mass & \multicolumn{2}{c}{$2.3~\mathrm{ng}$}\\
    $\xzpf$ & $\sqrt{\frac{\hbar}{2\meff\Om}}$ & Zero point fluctuations & \multicolumn{2}{c}{$1.8~\mathrm{fm}$}\\
    $\lambda_j$ & & Laser wavelength & $796.154~\mathrm{nm}$ & $796.750~\mathrm{nm}$ \\
    \multirow{2}{*}{$g_j$} & & Field-enhanced & \multirow{2}{*}{$2\pi\times67.0~\mathrm{kHz}$} & \multirow{2}{*}{$2\pi\times53.1~\mathrm{kHz}$} \\
    & & optomechanical coupling &\\
    $\kk{j}$ & & Cavity linewidth & $2\pi\times13.3~\mathrm{MHz}$ & $2\pi\times12.6~\mathrm{MHz}$\\
    $\Delta_j$ & & Laser-cavity detuning & $- 0.22\,\kk{A}$ & $- 0.20\,\kk{B}$\\
    $\eta^c_j$ & & Cavity overcoupling & $95\%$ & $95\%$ \\
    $\eta_j$ & & Detection efficiency & $60\%$ & $77\%$\\   
      &   &   & &  \\\hline    
    \multirow{2}{*}{$\Gamma^\mathrm{meas}_j$} & \multirow{2}{*}{$\eta_j\eta^\mathrm{c}_j\,4g_j^2/\kk{j}$} & \multirow{2}{*}{Measurement rate} & \multirow{2}{*}{$2\pi\times~0.77\mathrm{kHz}$} & \multirow{2}{*}{$2\pi\times~0.65\mathrm{kHz}$} \\
      &   &   & &  \\    
    \multirow{2}{*}{$\Gamma_j^\mathrm{qba}$} & \multirow{2}{*}{$4g_j^2/\kk{j}$} & Measurement-induced& \multirow{2}{*}{$2\pi\times~1.35\mathrm{kHz}$} & \multirow{2}{*}{$2\pi\times~0.89\mathrm{kHz}$}\\
      &   & quantum backaction rate &  & \\
    \multirow{2}{*}{$\gamma$} & \multirow{2}{*}{$\Gm \left(\bnth+1/2\right) $} & \multirow{2}{*}{Thermal decoherence rate} & \multicolumn{2}{c}{\multirow{2}{*}{$2\pi\times202~\mathrm{Hz}$}}\\ \\
    \multirow{2}{*}{$\eta_\mathrm{meas}$} &\multirow{2}{*}{$\sum_j\frac{\Gamma^\mathrm{meas}_j}{\Gamma_j^\mathrm{qba}+\gamma}$} & \multirow{2}{*}{Total measurement efficiency} & \multicolumn{2}{c}{\multirow{2}{*}{$58~\%$}} \\
    &   &   & &  \\    
    \multirow{2}{*}{$\Gdec$} & \multirow{2}{*}{$\Gamma_A^\mathrm{qba}+\Gamma_B^\mathrm{qba}+\gamma$}  & \multirow{2}{*}{Total decoherence rate}  & \multicolumn{2}{c}{\multirow{2}{*}{$2\pi\times2.44~\mathrm{kHz}$}}\\ \\ \hline
    $\theta_j$  &   & Homodyne detection angle  & &  \\ $\Theta$  & $(\theta_A + \theta_B)/2$  & joint homodynes angle  & &  \\   
    $\hat{X}_j^{\theta_j}$ &  & Quadrature of optical mode  & &  \\
    $\hat{X}_\pm$ & $\hat{X}_A^{\theta_A}\pm\hat{X}_B^{\theta_B}$  & EPR amplitude quadratures  & &  \\
    $\hat{Y}_\pm$ & $\hat{Y}_A^{\theta_A} \pm \hat{Y}_B^{\theta_B}$  & EPR phase quadratures  & &  \\
    $\nu_-$  &   & Minimum symplectic eigenvalue  & &  \\
    \multirow{2}{*}{$S_{\hat{A}\hat{B}}(\Og)$} & \multirow{2}{*}{$\int dt e^{\imath\Og t}\langle\hat{A}(t)\hat{B}(0)\rangle$}  & \multirow{2}{*}{Power spectral density}  & &  \\
    \\
    \multirow{2}{*}{$\overline{S}_{\hat{A}\hat{B}}(\Og)$} & \multirow{2}{*}{$\frac{S_{\hat{A}\hat{B}}(\Og)+S_{\hat{B}\hat{A}}(-\Og)}{2}$}  & \multirow{2}{*}{Symmetrized PSD}  & &  \\
    \\ \hline    
\end{tabular}
\caption{Parameters and definitions.}
\label{tab:parameters}
\end{table}

\end{document}